\documentstyle[aps,prl,preprint,floats,epsfig]{revtex}

\begin{document}

%\draft

% Comment out to get double spacing.
\tighten

%%%%%%%%%%%%%%%%%%%%%%%%%%%%%%%%%%%%%%%%%%%%%%%%%%%%%%%%%%%%%%%%%%%%%%%%%%%
\title{ \bf 
Measurements of $B \rightarrow D_s^{(*)+} D^{*(*)}$ Branching Fractions
}  

% Insert author list file here
% \input /home/axp/cleoac/AuthorList/Lists/0006/JOURNALS/clns-00-1683_prd.tex

\author{
S.~Ahmed, M.~S.~Alam, S.~B.~Athar, L.~Jian, L.~Ling, M.~Saleem,
S.~Timm, and F.~Wappler}
\address{
State University of New York at Albany, Albany, New York 12222}
\author{
A.~Anastassov, J.~E.~Duboscq, E.~Eckhart, K.~K.~Gan, C.~Gwon,
T.~Hart, K.~Honscheid, D.~Hufnagel, H.~Kagan, R.~Kass,
T.~K.~Pedlar, H.~Schwarthoff, J.~B.~Thayer, E.~von~Toerne,
and M.~M.~Zoeller}
\address{
Ohio State University, Columbus, Ohio 43210}
\author{
S.~J.~Richichi, H.~Severini, P.~Skubic, and A.~Undrus}
\address{
University of Oklahoma, Norman, Oklahoma 73019}
\author{
S.~Chen, J.~Fast, J.~W.~Hinson, J.~Lee, D.~H.~Miller,
E.~I.~Shibata, I.~P.~J.~Shipsey, and V.~Pavlunin}
\address{
Purdue University, West Lafayette, Indiana 47907}
\author{
D.~Cronin-Hennessy, A.L.~Lyon, and E.~H.~Thorndike}
\address{
University of Rochester, Rochester, New York 14627}
\author{
C.~P.~Jessop, H.~Marsiske, M.~L.~Perl, V.~Savinov, and X.~Zhou}
\address{
Stanford Linear Accelerator Center, Stanford University, Stanford,
California 94309}
\author{
T.~E.~Coan, V.~Fadeyev, Y.~Maravin, I.~Narsky, R.~Stroynowski,
J.~Ye, and T.~Wlodek}
\address{
Southern Methodist University, Dallas, Texas 75275}
\author{
M.~Artuso, R.~Ayad, C.~Boulahouache, K.~Bukin, E.~Dambasuren,
S.~Karamov, G.~Majumder, G.~C.~Moneti, R.~Mountain, S.~Schuh,
T.~Skwarnicki, S.~Stone, G.~Viehhauser, J.C.~Wang, A.~Wolf,
and J.~Wu}
\address{
Syracuse University, Syracuse, New York 13244}
\author{
S.~Kopp}
\address{
University of Texas, Austin, TX  78712}
\author{
A.~H.~Mahmood}
\address{
University of Texas - Pan American, Edinburg, TX 78539}
\author{
S.~E.~Csorna, I.~Danko, K.~W.~McLean, Sz.~M\'arka, and Z.~Xu}
\address{
Vanderbilt University, Nashville, Tennessee 37235}
\author{
R.~Godang, K.~Kinoshita,%
\thanks{Permanent address: University of Cincinnati, Cincinnati, OH 45221}
I.~C.~Lai, and S.~Schrenk}
\address{
Virginia Polytechnic Institute and State University,
Blacksburg, Virginia 24061}
\author{
G.~Bonvicini, D.~Cinabro, S.~McGee, L.~P.~Perera, and G.~J.~Zhou}
\address{
Wayne State University, Detroit, Michigan 48202}
\author{
E.~Lipeles, S.~P.~Pappas, M.~Schmidtler, A.~Shapiro, W.~M.~Sun,
A.~J.~Weinstein, and F.~W\"{u}rthwein%
\thanks{Permanent address: Massachusetts Institute of Technology, Cambridge, MA 02139.}}
\address{
California Institute of Technology, Pasadena, California 91125}
\author{
D.~E.~Jaffe, G.~Masek, H.~P.~Paar, E.~M.~Potter, S.~Prell,
and V.~Sharma}
\address{
University of California, San Diego, La Jolla, California 92093}
\author{
D.~M.~Asner, A.~Eppich, T.~S.~Hill, R.~J.~Morrison, H.~N.~Nelson,
J.~D.~Richman, and M.~S.~Witherell}
\address{
University of California, Santa Barbara, California 93106}
\author{
R.~A.~Briere  and G.~P.~Chen}
\address{
Carnegie Mellon University, Pittsburgh, Pennsylvania 15213}
\author{
B.~H.~Behrens, W.~T.~Ford, A.~Gritsan, J.~Roy, and J.~G.~Smith}
\address{
University of Colorado, Boulder, Colorado 80309-0390}
\newpage 
\author{
J.~P.~Alexander, R.~Baker, C.~Bebek, B.~E.~Berger, K.~Berkelman,
F.~Blanc, V.~Boisvert, D.~G.~Cassel, M.~Dickson, P.~S.~Drell,
K.~M.~Ecklund, R.~Ehrlich, A.~D.~Foland, P.~Gaidarev, L.~Gibbons,
B.~Gittelman, S.~W.~Gray, D.~L.~Hartill, B.~K.~Heltsley,
P.~I.~Hopman, C.~D.~Jones, D.~L.~Kreinick, M.~Lohner,
A.~Magerkurth, T.~O.~Meyer, N.~B.~Mistry, E.~Nordberg,
J.~R.~Patterson, D.~Peterson, D.~Riley, J.~G.~Thayer, D.~Urner,
B.~Valant-Spaight, and A.~Warburton}
\address{
Cornell University, Ithaca, New York 14853}
\author{
P.~Avery, C.~Prescott, A.~I.~Rubiera, J.~Yelton, and J.~Zheng}
\address{
University of Florida, Gainesville, Florida 32611}
\author{
G.~Brandenburg, A.~Ershov, Y.~S.~Gao, D.~Y.-J.~Kim, and R.~Wilson}
\address{
Harvard University, Cambridge, Massachusetts 02138}
\author{
T.~E.~Browder, Y.~Li, J.~L.~Rodriguez, and H.~Yamamoto}
\address{
University of Hawaii at Manoa, Honolulu, Hawaii 96822}
\author{
T.~Bergfeld, B.~I.~Eisenstein, J.~Ernst, G.~E.~Gladding,
G.~D.~Gollin, R.~M.~Hans, E.~Johnson, I.~Karliner, M.~A.~Marsh,
M.~Palmer, C.~Plager, C.~Sedlack, M.~Selen, J.~J.~Thaler,
and J.~Williams}
\address{
University of Illinois, Urbana-Champaign, Illinois 61801}
\author{
K.~W.~Edwards}
\address{
Carleton University, Ottawa, Ontario, Canada K1S 5B6 \\
and the Institute of Particle Physics, Canada}
\author{
R.~Janicek  and P.~M.~Patel}
\address{
McGill University, Montr\'eal, Qu\'ebec, Canada H3A 2T8 \\
and the Institute of Particle Physics, Canada}
\author{
A.~J.~Sadoff}
\address{
Ithaca College, Ithaca, New York 14850}
\author{
R.~Ammar, A.~Bean, D.~Besson, R.~Davis, N.~Kwak, and X.~Zhao}
\address{
University of Kansas, Lawrence, Kansas 66045}
\author{
S.~Anderson, V.~V.~Frolov, Y.~Kubota, S.~J.~Lee, R.~Mahapatra,
J.~J.~O'Neill, R.~Poling, T.~Riehle, A.~Smith, C.~J.~Stepaniak,
and J.~Urheim}
\address{
University of Minnesota, Minneapolis, Minnesota 55455}
 
\author{(CLEO Collaboration)} 

\address{} \date{\today} \maketitle

\begin{abstract} 
% Insert abstract here.
This article describes improved measurements by CLEO of the
$B^0 \rightarrow D_s^+ D^{*-}$ and $B^0 \rightarrow D_s^{*+} D^{*-}$ branching
fractions, and first evidence for the decay
$B^+ \rightarrow D_s^{(*)+} \bar{D}^{**0}$, where $\bar{D}^{**0}$
represents the sum of the $\bar{D}_1(2420)^0$, $\bar{D}_2^*(2460)^0$, and
$\bar{D}_1(j=1/2)^0$ $L=1$ charm meson states.  Also reported is the first
measurement of the $D_s^{*+}$ polarization in the decay
$B^0 \rightarrow D_s^{*+} D^{*-}$.  A partial reconstruction technique,
employing only the fully reconstructed $D_s^+$ and slow pion $\pi_s^-$ from the
$D^{*-} \rightarrow \bar{D}^0 \pi^-_s$ decay, enhances sensitivity.
The observed branching fractions are
$ { \mathcal B} (B^0 \rightarrow D_s^+ D^{*-})  =
    (1.10 \pm 0.18 \pm 0.10 \pm 0.28)\%$ ,
$ { \mathcal B} (B^0 \rightarrow D_s^{*+} D^{*-})  =
    (1.82 \pm 0.37 \pm 0.24 \pm 0.46)\%$ , and
$ { \mathcal B} (B^+ \rightarrow D_s^{(*)+} \bar{D}^{**0})  =
    (2.73 \pm 0.78 \pm 0.48 \pm 0.68)\%$,
where the first error is statistical, the second systematic, and the
third is the uncertainty in the $D_s^+ \rightarrow \phi \pi^+$ branching
fraction.  The measured $D_s^{*+}$ longitudinal polarization, $\Gamma_L/\Gamma
= (50.6 \pm 13.9 \pm 3.6)\%$, is consistent with the factorization prediction
of 54\%.

\end{abstract}

%%%%%%%%%%%%%%%%%%%%%%%%%%%%%%%%%%%%%%%%%%%%%%%%%%%%%%%%%%%%%%%%%%%%%%%%%%%
% Begin main body of text.

\section{Introduction}
\label{intro}

\begin{figure}[htbp] \begin{center}
\epsfig{figure=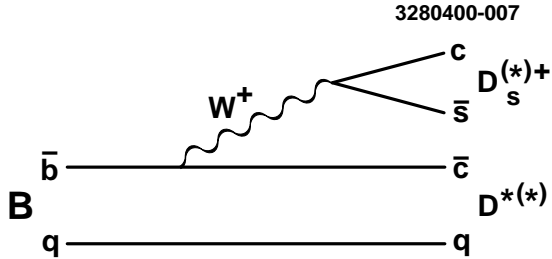, width=3.00in}
\caption{ The spectator diagram for $B \rightarrow D_s^{(*)+} D^{*(*)}$
          decay.
         }
\label{feynman} 
\end{center} \end{figure}

Measurements of weak decays of $B$ mesons are fundamental to testing and 
understanding the standard model. 
Previous measurements of the inclusive $B \rightarrow D_s^+ X$ branching
fraction report a value of $(12.1 \pm 1.0 \pm 3.0)\%$.  The first
error is the combined statistical and systematic uncertainties, and the second
is due to the uncertainty in the $D_s^+ \rightarrow \phi \pi^+$ branching 
fraction.  This is significantly larger than the sum of $D_s^+$ production
from exclusive $b \rightarrow c \bar{c} s$ modes observed
to date\cite{B-DsX}.  These exclusive modes, of the form
$B \rightarrow D_s^+ D$, $B \rightarrow D_s^{*+} D$,
$B \rightarrow D_s^+ D^{*}$, and $B \rightarrow D_s^{*+} D^{*}$, sum to
$(6.6 \pm 1.3 \pm 1.7)\%$ for the $B^+$ case and $(4.8 \pm 1.0 \pm 1.2)\%$
for the $B^0$.  This yields a deficit of $(5.5 \pm 1.6)\%$ for the $B^+$ 
and $(7.3 \pm 1.4)\%$ for the $B^0$, where
the $D_s^+ \rightarrow \phi \pi^+$ branching fraction uncertainty does
not affect this difference\cite{B-DsX}. 
This article reports new measurements of $B \rightarrow D_s^{(*)+} D^{*(*)}$
decays from CLEO.\footnote{Reference to a specific state or decay includes
the charge-conjugate state or decayThe notation 
$D_s^{(*)+}$ in this context means either a $D_s^+$ or a $D_s^{*+}$,
$D^{*(*)}$ denotes the sum of $D^*$ and $D^{**}$, and $D^{**}$ denotes
the sum of the charged $D^{**+}$ and neutral $\bar{D}^{**0}$ states,  
the specifics of which are discussed in Section~\ref{d_dub}.
In shortened form,
$D_s D^*$ denotes $D_s^+ D^{*-}$, $D_s^* D^*$ denotes
$D_s^{*+} D^{*-}$, and $D_s^{(*)} D^{**}$ denotes the sum of
$D_s^{(*)+} D^{**-}$ and $D_s^{(*)+} \bar{D}^{**0}$. }
% First evidence is offered for the decay 
% $B^+ \rightarrow D_s^{(*)+} \bar{D}^{**0}$, which may bridge a substantial
% portion of the inclusive and exclusive rate difference, where 
% $\bar{D}^{**0}$ denotes the sum of the $\bar{D}_1(2420)^0$,
% $\bar{D}_2^*(2460)^0$, and $\bar{D}_1(j=1/2)^0$ $L=1$ charm meson states.
First evidence is offered for the decay $B^+ \rightarrow D_s^{(*)+}
\bar{D}^{**0}$,  where $\bar{D}^{**0}$ denotes the sum of the
$\bar{D}_1(2420)^0$, $\bar{D}_2^*(2460)^0$, and $\bar{D}_1(j=1/2)^0$ $L=1$ 
charm meson states.  This decay mode may bridge a substantial portion of the
inclusive and exclusive rate difference.
Also reported are improved measurements of the modes
$B^0 \rightarrow D_s^+ D^{*-}$ and $B^0 \rightarrow D_s^{*+} D^{*-}$.
These decays occur predominantly via the spectator diagram of
Figure~\ref{feynman}; the $W^+$ decays into a $D_s^+$ or $D_s^{*+}$ meson, 
and the charm anti-quark and spectator quark hadronize as either a 
$D^{*}$ or $D^{**}$ meson.
% \begin{figure}[htbp]
%   \epsfysize=9cm    % smaller for prl double column
% \begin{center}
% \epsfysize=4.50cm % single column mode for clns (15cm)
% \epsfbox{3280400-007.ps}
% \caption{\label{feynman}
%        The spectator diagram for $B \rightarrow D_s^{(*)+} D^{*(*)}$
%        decay.
%        }    
% \end{center}
% \end{figure}

Additionally, this Article presents the first measurement of $D_s^{*+}$
polarization for the mode $B^0 \rightarrow D_s^{*+} D^{*-}$, providing an
effective test of the factorization assumption in $B \rightarrow D^{*-} X$
decays with high $q^2$, where $q^2 = M_X^2$, and $X$ is a vector meson.
Factorization assumes the lack of final state interactions
between the products of hadronic $B$ decays, and has successfully
predicted the vector-vector polarization of the low $q^2$ mode
$B \rightarrow D^{*-}
\rho$\cite{Theory1,Theory2,Theory3,Theory4,Theory5}. 
It is possible that the factorization assumption of no final state interactions
may be simplistic and inapplicable to modes of higher $q^2$ such as 
$B^0 \rightarrow D_s^{*+} D^{*-}$; however, the results
presented here are consistent with the factorization prediction.  

Previous measurements of $B^0 \rightarrow D_s^+ D^{*-}$ and
$B^0 \rightarrow D_s^{*+} D^{*-}$ at CLEO and ARGUS made use of the
full reconstruction technique\cite{B-DsX}\cite{B-DsX2}, which requires
reconstruction of all particles in the final state.  The most recent
CLEO results using full reconstruction reported relatively small event yields
of $18.4 \pm 4.5$ and $17.7 \pm 4.4$ in the $D_s^+ D^{*-}$ and
$D_s^{*+} D^{*-}$ channels, respectively.
Following these, a partial reconstruction technique was developed
that required only some of the $B^0 \rightarrow D_s^{*+} D^{*-}$ final state
particles, reporting an increased sample size of
$76 \pm 11$ events\cite{Ds-PhP}.

This analysis employs a more refined partial reconstruction technique, using
only the $D_s^+$ and the soft pion $\pi_s^-$ from the
$D^{*-} \rightarrow \bar{D}^0 \pi_s^-$ decay, thereby increasing the statistics
over full reconstruction by a factor between five and eight, depending on mode.
The analysis is sensitive to any $B \rightarrow D_s^+ D^{*-} X$ final state,
such as $B \rightarrow D_s^{(*)+} D^{**}$, when $D^{**} \rightarrow D^{*-} \pi$.
The method is based on techniques developed by CLEO for improved measurement of
$D_s^+ \rightarrow \phi \pi^+$\cite{Ds-PhP} and 
$B \rightarrow D^{*} \pi$\cite{B-DPi}. 

After a short description of the detector and the criteria used for selecting
charged particle candidates in Section~\ref{event_sel}, the $D_s^+$ and 
$\pi_s^-$ reconstruction is described in Section~\ref{ds_rec}.  In 
Section~\ref{1d_discuss} the partial reconstruction technique is developed
for separating the combined $D_s^{(*)} D^{*(*)}$ signal from background.  Once
the background levels have been determined, in Section~\ref{2d_discuss} a
two-dimensional parameter space is defined and used to separate the
individual $D_s D^*$, $D_s^* D^*$, and $D_s^{(*)} D^{**}$ signals, followed
by a review
of systematic errors in Section~\ref{sys_discuss}.  The polarization of
$D_s^* D^*$ production is measured and compared with the factorization
prediction in Section~\ref{factorization}, and the results summarized and
discussed in the final section.

% Section I	-- Introduction
% Section II    -- Event Selection
% Section III   -- $D^+_s$ and Slow $\pi^-_s$ Reconstruction 
% Section IV    -- Separation of $D_s^{(*)+} D^{*(*)}$ Signal from Background
% Section V     -- Separation of $D_s D^*$, $D_s^* D^*$, and $D_s^{(*)} D^{**}$
%                  Signals.  Measurement of $D_s^* D_s^*$ Polarization. 
% Section VI    -- Systematic Errors
% Section VII   -- Factorization and Polarization Prediction
% Section VIII  -- Conclusions 

\section{EVENT SELECTION}
\label{event_sel}

The data used in this analysis were collected at the Cornell Electron Storage
Ring (CESR) between 1990 and 1995, and consist of hadronic events produced
in $e^+ e^-$ annihilations.  The integrated luminosity of this data sample
is $3.14 \pm 0.06$ ${\rm fb^{-1}}$ collected at the $\Upsilon(4S)$ resonance
(referred to as on-resonance data), and $1.69 \pm 0.03$ ${\rm fb^{-1}}$ from
a center-of-mass energy just below the threshold for producing $B\bar{B}$
mesons (referred to as off-resonance or continuum data).  The on-resonance
data corresponds to $(3.36 \pm 0.06) \times 10^6$ $B\bar{B}$ pairs.

The CLEO II detector is used to measure both neutral and charged particles
with excellent resolution and efficiency\cite{CLEOII}.  Hadronic events are 
selected by requiring a minimum of three charged tracks, a total visible
energy greater than 15\% of the center-of-mass energy (this reduces 
contamination from two-photon interactions and beam-gas events), and a 
primary vertex within $\pm 5$ cm in the $z$ direction and $\pm 2$ cm 
in the $r$-$\phi$ plane of the beam centroid.

Charged tracks are required to be of good quality and consistent with
the primary vertex in both the $r$-$\phi$ and $r$-$z$ planes.  Tracks
must also have $dE/dx$ and time-of-flight information consistent with
their pion or kaon hypotheses, when such information exists and is of
good quality.

Apart from the visible energy criterion, neutral particles were not used in
this analysis.  

A {\sc geant}\cite{Geant} based Monte Carlo simulation was used to generate
large samples of the individual $D_s^{(*)} D^{*(*)}$ signal modes from
$\Upsilon(4S) \rightarrow B \bar{B}$ decays, and model their interactions
with the CLEO detector.  These samples were then processed
in the same manner as the data.  Further discussion of the simulation is given
in the treatment of systematic errors.

\section{$D^+_s$ AND SLOW $\pi^-_s$ RECONSTRUCTION}
\label{ds_rec}

\begin{figure}[htbp] \begin{center}
\epsfig{figure=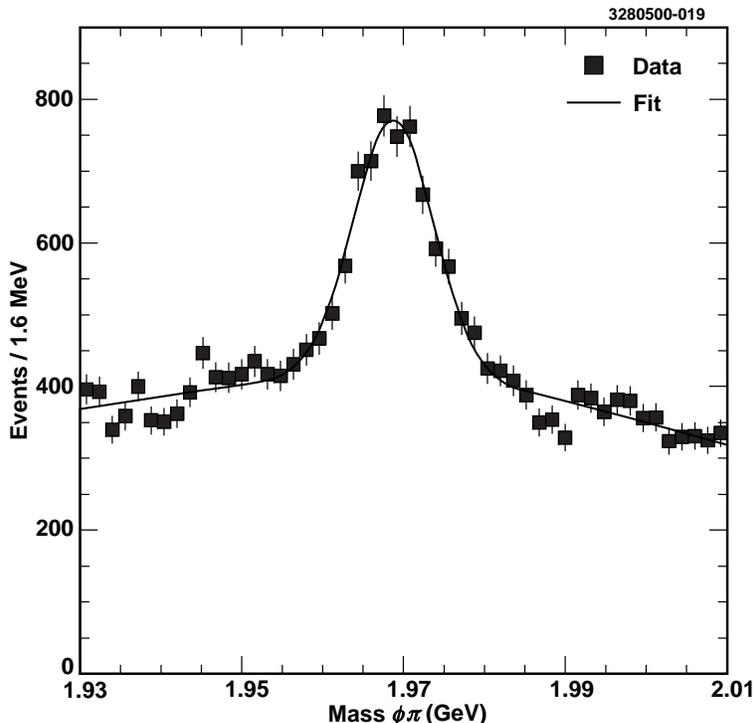, width=4.00in}
\caption{ The $\phi \pi^+$ mass spectrum for the on-resonance data.
          The $\phi \pi^+$ mass is further required to be within 12 MeV
          of the $D_s^+$ mass. 
          }
\label{ds_mass} 
\end{center} \end{figure}

The $D_s^+$ is reconstructed through the $D_s^+ \rightarrow \phi \pi^+$,
$\phi \rightarrow K^+ K^-$ decay channel, which has a signal-to-background
ratio nearly two times higher than the next cleanest $D_s^+$ decay
mode\cite{Ds-PhP}.  Fast $\pi^+$/$K^+$ tracks ($p \ge 200 \ {\rm MeV}/c$)
must originate within $\pm 5$ cm in the $z$ direction and $\pm 5$ mm in the
$r$-$\phi$ plane of the beam centroid.  For slow $\pi^+$/$K^+$ tracks
($p \le 200 \ {\rm MeV}/c$) the $z$ requirement is loosened to within
$\pm 20$ cm.  The $K^+ K^-$ invariant mass is required to be within 9 MeV of
the $\phi$ mass.  Two angles are used in suppressing background. The first is
the $D_s^+$ decay angle $\theta_D$, which is the angle between the $\phi$
direction in the $D_s^+$ rest frame and the $D_s^+$ boost direction.  Requiring
$\cos{\theta_D} \le 0.80$ eliminates a large combinatoric background peak near
$\cos{\theta_D} = 1$ resulting from the numerous low momentum pions, while the
signal is constant in $\cos{\theta_D}$.
The second angle is $\theta_H$, the $\phi$ decay angle between the
$K^+$ and $D_s^+$ direction in the $\phi$ rest frame.  Due to the $\phi$
helicity the signal follows a $\cos^2{\theta_H}$ distribution while the
background is constant in $\cos{\theta_H}$.  Requiring
$|\cos{\theta_H}| \ge 0.35$ removes 35\% of the background and retains 96\%
of signal.  
The resulting $\phi \pi^+$ invariant mass spectrum is shown in
Figure~\ref{ds_mass}, and the $\phi \pi^+$ mass is then required to be
within 12 MeV of the $D_s^+$ mass.
Finally, the
kinematics of $B \rightarrow D_s^{(*)+} D^{**}$ decays constrain the
magnitude of $D_s^+$ momentum to between 1250 MeV/$c$ and 1925 MeV/$c$, and
these requirements are imposed here.

The slow pion $\pi^-_s$ from the $D^{*-}$ must have charge opposite to
the $D_s^+$ and originate within $\pm 5$ mm in the $r$-$\phi$ plane of the
primary vertex.  No $z$
requirement is placed on the $\pi^-_s$, but it must have a momentum greater
than 50 MeV/$c$ and less than 210 MeV/$c$.

\section{SEPARATION OF $D_s^{(*)+} D^{*(*)}$ SIGNAL FROM BACKGROUND}
\label{1d_discuss}

\subsection{Two-Body $B$ decays to $D_s^+ D^{*-}$ Final States}
\label{d_dub}

At the CLEO II experiment, $e^+ e^-$ collisions can create an $\Upsilon(4S)$
resonance, which decays to a pair of $B$ mesons.  The $B$'s are produced
nearly at rest ($\beta = 0.0646$) and, for the decay chain
$\Upsilon(4S) \rightarrow B^0 \bar{B}^0$, $B^0 \rightarrow D_s^+ D^{*-}$,
and $D^{*-} \rightarrow \bar{D^0} \pi_s^-$, the $D_s^+$ and soft pion $\pi^-_s$
are nearly back-to-back in the lab frame because of the small $5.83 \pm 0.03$
MeV energy release in the $D^{*-} \rightarrow \bar{D^0} \pi_s^-$
transition.  By making use of their relative direction, as well as the beam
energy and kinematic constraints of the decay, the $D_s^+$ and the $\pi_s^-$
allow reconstruction of the $D_s^+ D^{*-}$ final state.  

Other two-body $B$ decays leading to $D_s^+ D^{*-}$ final states, with
strong $(D_s^+, \pi_s^-)$ correlations, are summarized in
Table~\ref{detected_modes}.
These are modes producing a $D_s^{*+}$ that decays to $D_s^+ \gamma$
or $D_s^+ \pi^0$, or producing a $D^{**}$ that decays to $D^{*-} \pi$.
It should be noted that this method is not sensitive to
$B \rightarrow D_s^{**} D^{*-}$,
as the $D_s^{**}$ decays predominantly to $D K$ and no
$D_s^{**} \rightarrow D_s X$ decays have been observed\cite{PDG}.  Other relevant modes,
such as three-body $B$ decays of the form $B \rightarrow D_s^{(*)+} D^{*-} \pi$,
are treated in the discussion of systematic errors.  
\begin{table}[htb]
\caption{$D_s^+ D^{*-}$ final states from two-body $B$ decays. }
\begin{center}
\begin{tabular}{ll}
$B^0$ Decays    &  $B^+$ Decays \\ \hline
$B^0 \rightarrow D_s^+ D^{*-}$ &  \\
$B^0 \rightarrow D_s^{*+} D^{*-}$ &  \\
 \hspace{0.32cm} (where $D_s^{*+} \rightarrow$    & \\
 \hspace{0.32cm} $D_s^+ \gamma$ / $D_s^+ \pi^0$)  & \\
$B^0 \rightarrow D_s^+ D^{**-}$ & $B^+ \rightarrow D_s^+ \bar{D}^{**0}$ \\
 \hspace{0.32cm} ($D^{**-} \rightarrow D^{*-} \pi^0$) &
 \hspace{0.32cm} ($\bar{D}^{**0} \rightarrow D^{*-} \pi^+$) \\
$B^0 \rightarrow D_s^{*+} D^{**-}$ & $B^+ \rightarrow D_s^{*+} \bar{D}^{**0}$\\
 \hspace{0.32cm} ($D^{**-} \rightarrow D^{*-} \pi^0$ &
 \hspace{0.32cm} ($\bar{D}^{**0} \rightarrow D^{*-} \pi^+$ \\
 \hspace{0.32cm} and $D_s^{*+} \rightarrow$ &
 \hspace{0.32cm} and $D_s^{*+} \rightarrow$ \\
 \hspace{0.32cm} $D_s^+ \gamma$ / $D_s^+ \pi^0$) &
 \hspace{0.32cm} $D_s^+ \gamma$ / $D_s^+ \pi^0$) \\ 
\end{tabular}
\end{center}
\label{detected_modes}
\end{table} 

\subsection{$D^{**}$ Properties and $B \rightarrow D_s^{(*)+} D^{**}$ Decays}

In this measurement several different $D^{**}$ states contribute to the 
$B \rightarrow D_s^{(*)+} D^{*(*)}$ decays.
The relevant $D^{**}$ characteristics are summarized here, beginning with the
neutral $D^{**0}$ which, as an $L=1$ charm meson, represents four distinct
quantum states.  Two of these states, the $D_1(2420)^0$ and $D_2^*(2420)^0$,
have been characterized by experiment as relatively narrow
resonances\cite{D-Dub1,D-Dub3}.  The two other states, the $D_1(j=1/2)^0$ and
$D_0(j=1/2)^0$, are expected to be much broader\cite{Theory3}. 
A preliminary first observation of the $D_1(j=1/2)^0$, confirming its
broadness, was recently reported by CLEO\cite{Wide-D}.  Although the
$D_0(j=1/2)^0$ remains experimentally undetected, conservation of parity and
angular momentum forbids its decay to $D^{*-} \pi^+$, so it does not 
contribute to this measurement.  Table~\ref{d_dub_props} gives the masses,
widths, $J^P$ and allowed decays of the neutral 
$D^{**}$'s\cite{D-Dub1,D-Dub3,Wide-D}.  
\begin{table}[htb]
\caption{$D^{**0}$ Properties}
\begin{center}
\begin{tabular}{lllll}
State & $J^P$ & Mass (MeV) & Width (MeV) & Allowed \\
      &       &            &             & Decays  \\ 
\hline
$D^*_0(j=1/2)^0$ &$0^+$& \it Not Yet Observed & ---             & $D \, \pi$ \\
$D_1(2420)^0$    &$1^+$& $2422.0 \pm 1.8$ & $18.9^{+4.6}_{-3.5}$ & $D^* \pi$ \\
$D_1(j=1/2)^0$   &$1^+$& $2461^{+42}_{-35}$& $290^{+104}_{-83}$& $
D^* \pi$ \\
$D^*_2(2460)^0$  &$2^+$& $2458.9 \pm 2.0$ & $23 \pm 5$ & $D \, \pi, D^* \pi$ \\
\end{tabular}
\end{center}
\label{d_dub_props}
\end{table} 

In accordance with current experimental limits, the masses and decay widths of
the charged $D_1(j=1/2)^-$, $D_1(2420)^-$, and $D_2^*(2460)^-$ are assumed
identical to their corresponding neutral $D^{**0}$ counterparts\cite{D-Dub2}.   
Like the $D_0(j=1/2)^0$, the $D_0(j=1/2)^-$ does not decay to
$D^{*-} \pi$. 

Throughout this Article $D^{**+}$ denotes the sum of the charged $D_1(2420)^+$,
$D_2^*(2460)^+$, and $D_1(j=1/2)^+$ states, while $D^{**0}$ denotes
the sum of the neutral $D_1(2420)^0$, $D_2^*(2460)^0$, and $D_1(j=1/2)^0$,
and $D^{**}$ denotes the sum of the three $D^{**+}$ and three $D^{**0}$
states.  

Conservation of isospin and angular momentum predicts the branching
fractions for the $J=1$ charged and neutral $D^{**} \rightarrow D^{*-} \pi$ 
decays.  Heavy quark effective chiral perturbation theory evaluates the 
branching fractions for the $J=2$ case\cite{Theory2}.
\begin{eqnarray}
{\mathcal B}(D_1(j=1/2)^-        \rightarrow D^{*-} \pi^0) & = & 1/3, \\
{\mathcal B}(D_1(2420)^-         \rightarrow D^{*-} \pi^0) & = & 1/3, \\
{\mathcal B}(D^*_2(2460)^-       \rightarrow D^{*-} \pi^0) & = & 1/10,\\
{\mathcal B}(\bar{D}_1(j=1/2)^0  \rightarrow D^{*-} \pi^+) & = & 2/3, \\
{\mathcal B}(\bar{D}_1(2420)^0   \rightarrow D^{*-} \pi^+) & = & 2/3, \\
{\mathcal B}(\bar{D}^*_2(2460)^0 \rightarrow D^{*-} \pi^+) & = & 1/5. 
\end{eqnarray}
These branching fractions are assumed throughout.  Applying conservation of
isospin to the spectator decay of Figure~\ref{feynman}, it is assumed also
that the $B^+ \rightarrow D_s^{(*)+} \bar{D}^{**0}$ and
$B^0 \rightarrow D_s^{(*)+} D^{**-}$ production rates are equal:
\begin{equation}
\Gamma(B^+ \rightarrow D_s^{(*)+} \bar{D}^{**0}) = 
 \Gamma(B^0 \rightarrow D_s^{(*)+} D^{**-}). 
\end{equation}
This equality is assumed throughout.  

In the Monte Carlo simulation, the three relevant $D^{**0}$ mass states
produce nearly
identical slow pion $\pi_s^-$ momentum distributions, resulting in signatures
that are virtually indistinguishable by means of this partial reconstruction
technique.  For this reason the relative production ratios of $D_1(j=1/2)^0$,
$D_1(2420)^0$, and $D_s^*(2420)^0$ cannot be measured by this analysis, but
are rather taken from previous experimental results\cite{Wide-D,D-Dub2}.  
Similarly, it was not possible to separate the $B \rightarrow D_s^+ D^{**}$
from the $B \rightarrow D_s^{*+} D^{**}$ modes, and the ratio of the branching
fractions of these two decays
must be assumed.  The consequences of both assumptions are
treated in the discussion of systematic errors.

%
% OK -- WHAT'S MISSING??
%
% (1)  Full Introduction.                               [STARTED]
% (2)  D** Discussion.                                  [DONE]
% (3)  Detailed explanation of partial reconstruction.  [DONE]
% (4)  Definition of alpha, alpha_0, cones.             [DONE]
% (5)  Definition of cos(theta_pi).                     [DONE]
% (6)  2-D plots of p_Ds vs cos(theta_pi).              [DONE]
% (7)  Projections of 2-D fit Results.                  [DONE]
% (8)  How do we want to deal with the "No D**" fit?    [DONE]
% (9)  Polished Systematics.                            [DONE]
% (10) Polished Conclusions.                            [STARTED]
% 

\subsection{Partial Reconstruction Kinematics} 

\begin{figure}[htbp] \begin{center}
\epsfig{figure=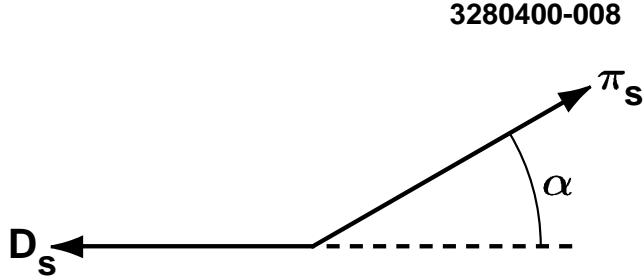, width=3.50in}
\caption{ Definition of $\alpha$: the angle between the reverse direction of
          the measured $D_s^+$  and measured $\pi_s^-$.  
          }
\label{alpha} 
\end{center} \end{figure}

In the decays $B \rightarrow D_s^{(*)+} D^{*(*)}$, $D^{*-} \rightarrow
\bar{D}^0 \pi_s^-$, the $D_s^+$ and $\pi_s^-$ are produced nearly back-to-back,
and the angle $\alpha$ between the reverse $D_s^+$ direction and $\pi_s^-$, 
shown in Figure~\ref{alpha}, will be small.  
For the $D_s D^*$ and $D_s^* D^*$ signals, $\alpha$ ranges between $0^\circ$
and $30^\circ$, with most probable values at 11$^\circ$ and $12^\circ$,
respectively. 
For $D_s^{(*)} D^{**}$ signal, $\alpha$ ranges from $0^\circ$ to $50^\circ$,
with the most probable value at $21^\circ$. 
% This results in the sharp peaking
% displayed by the signal Monte Carlo distributions of Figure~\ref{alpha_mc},
% which show $\cos{\alpha}$ combined with $\cos{\alpha_0}$, where $\alpha_0$
% is to be defined shortly.  
In contrast to the signal, the background consists of uncorrelated
$D_s^+$ and $\pi_s^-$ pairs, for which $\alpha$ will be distributed
at random.  

It is possible to further constrain $\alpha$ from additional event information
that determines the allowed $D^{*-}$ directions.  There exist a total of eight
unknowns in the decay: the $\bar{D}^0$ three-momentum, the parent $D^{*-}$
three-momentum, and the two angles governing the $B^0$ direction.  The $B$
energy is equal to the CLEO beam energy. 
Requiring conservation of energy and momentum in the
$B^0 \rightarrow D_s^+ D^{*-}$ and $D^{*-} \rightarrow \bar{D}^0 \pi^-_s$
decays yields eight constraints, where the masses of the $B^0$, $D_s^+$,
$D^{*-}$, $\bar{D}^0$, and $\pi^-$ are assumed.  Solving for the unknowns
yields a pair of $D^{*-}$ solutions, due to a quadratic ambiguity in the
underlying algebra.  The procedure of this solution follows.

The $D^{*-}$ and $\bar{D}^0$ energies are determined from the measured
$D_s^+$ and $\pi_s^+$ energies:

\begin{equation}
E_{D^{*-}} = E_{Beam} - E_{D_s^+}, 
\end{equation} 

\begin{equation}
E_{\bar{D}^0} = E_{D^{*-}} - E_{\pi_s^+}. 
\end{equation}

\begin{figure}[htbp] \begin{center}
\epsfig{figure=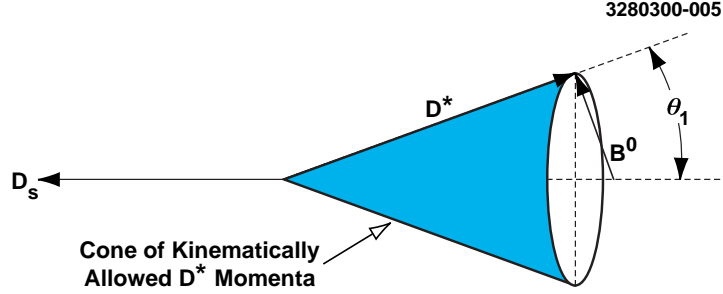, width=4.00in}
\caption{ Definition of $\theta_1$: the angle between the reverse $D_s^+$
          direction and the cone of allowed $D^{*-}$ directions.  
          }
\label{theta_1} 
\end{center} \end{figure}

The magnitude of $D^{*-}$ and $\bar{D}^0$ momenta follow from their
energies $p_{D^{*-}} = \sqrt{E_{D^{*-}}^2 - M_{D^{*-}}^2}$ and 
$p_{\bar{D}^0} = \sqrt{E_{\bar{D}^0}^2 - M_{\bar{D}^0}^2}$.
For the previously-assumed decay, kinematics constrain the $D^{*-}$
to a cone of allowed directions relative to the measured $D_s^+$.  The radius
of this cone, $\theta_1$, is shown in Figure~\ref{theta_1}, and represents
the angle between the reverse $D_s^+$ direction and inferred $D^{*-}$. 
Using the $D_s^+$ momentum magnitude, beam energy, and particle masses,
$\theta_1$ can be expressed in the lab frame as: 

\begin{equation}
\cos{\theta_1} = \frac{M^2_{B^0}-M^2_{D^*}-M^2_{D_s}}
                      {2 |\vec{p}_{D_s}| |\vec{p}_{D^*}| }
               - \frac{1}{\beta_{D_s} \beta_{D^*}}. 
\end{equation}

\begin{figure}[htbp] \begin{center}
\epsfig{figure=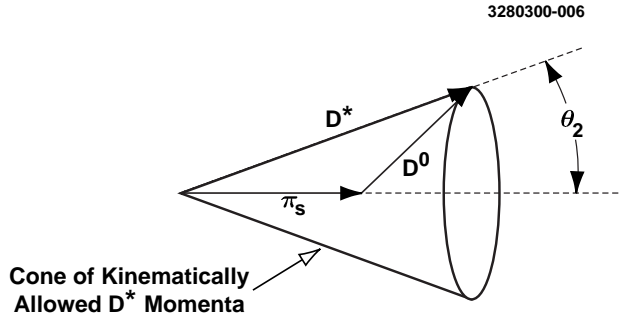, width=4.00in}
\caption{ Definition of $\theta_2$: the angle between the $\pi_s^-$ and cone
          of allowed $D^{*-}$ directions. 
          }
\label{theta_2} 
\end{center} \end{figure}

Kinematics also constrain the $D^{*-}$ to a cone of allowed values about the 
$\pi_s^-$ direction.  The radius of this cone is $\theta_2$, the angle between
the $\pi_s^-$ and inferred $D^{*-}$, defined in the lab frame and shown 
in Figure~\ref{theta_2}:  

\begin{equation}
\cos{\theta_2} = \frac{M^2_{D^0}-M^2_{D^*}-M^2_{\pi_s}}
                        {2 |\vec{p}_{\pi_s}| |\vec{p}_{D^*}| }
                 + \frac{1}{\beta_{\pi_s} \beta_{D^*}}. 
\end{equation}

\begin{figure}[htbp] \begin{center}
\epsfig{figure=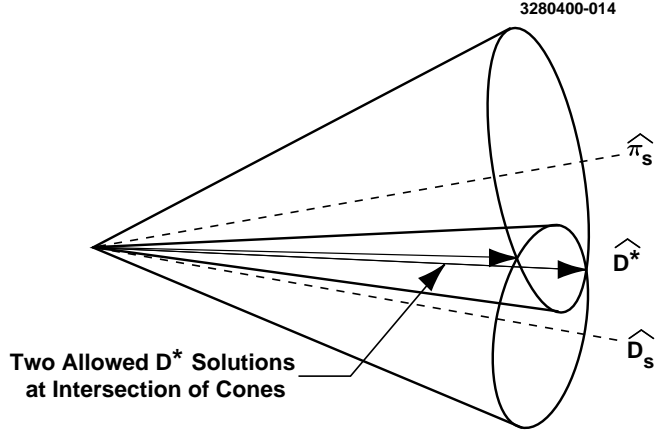, width=4.00in}
\caption{ Valid $D^{*-}$ solutions exist where the combined $\theta_1$ and 
          $\theta_2$ cones intersect.  There are generally two solutions, 
          resulting from a quadratic ambiguity in the underlying algebra.
          }
\label{comb_cone} 
\end{center} \end{figure}

Valid solutions for the $D^{*-}$ momentum exist at the intersection of the
cones defined by
$\theta_1$ and $\theta_2$, as shown in Figure~\ref{comb_cone}.  For
the two cones to intersect, the angle between the measured $D_s^+$ and
measured $\pi_s^-$---the angle previously defined as $\alpha$---must be
confined to a range bounded by the sum and difference of $\theta_1$ and
$\theta_2$:

\begin{equation}
| \theta_1 - \theta_2 | \le \alpha \le \theta_1 + \theta_2. 
\end{equation} 

\begin{figure}[htbp] \begin{center}
\epsfig{figure=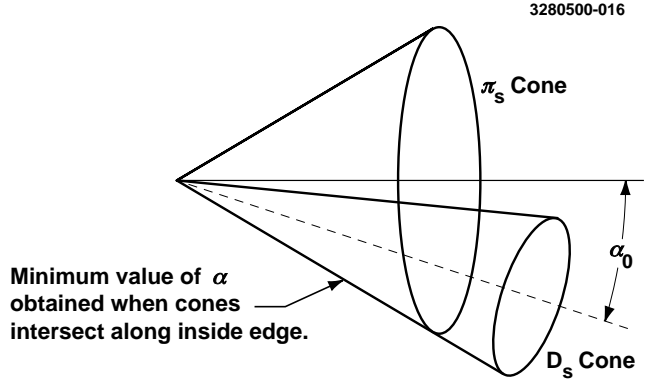, width=3.50in}
\caption{ $\alpha_0$ is the minimum value $\alpha$ can take for an event
          where the cones continue to intersect, and corresponds to the smaller
          cone grazing the inside edge of the larger.
          }
\label{alpha0} 
\end{center} \end{figure}

For $\alpha$ greater than the upper limit $\theta_1 + \theta_2$, the
smaller cone is entirely outside the larger one, preventing their intersection
and the existence of a kinematically valid $D^{*-}$ solution.  
For $\alpha$ less than the lower limit $| \theta_1 - \theta_2 |$, the
smaller cone is completely inside the larger, also preventing their 
intersection.  As shown in Figure~\ref{alpha0}, the lower limit occurs as
the smaller cone grazes the inside edge of the larger one, where this limit
is defined as $\alpha_0$: 
\begin{equation}
\alpha_0 \equiv | \theta_1 - \theta_2 |. 
\end{equation}
Since only one $\alpha$ and one $\alpha_0$ exist for a particular $(D_s^+,
\pi_s^-)$ pair, they are unaffected by the
quadratic ambiguity in the $D^{*-}$ solutions.  

\begin{figure}[htbp] \begin{center}
\epsfig{figure=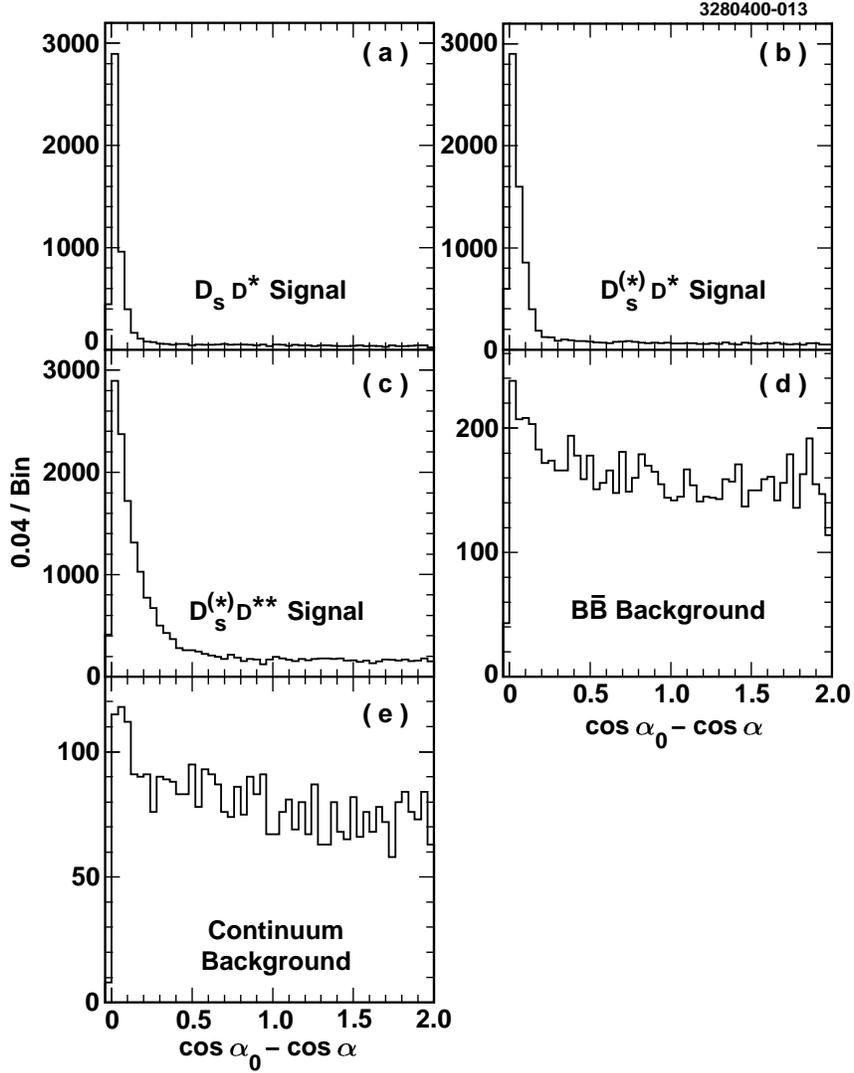, width=4.50in}
\caption{ Signal Monte Carlo and background distributions of the
          partial-reconstruction variable $\cos{\alpha_0}-\cos{\alpha}$.  
          Shown are: (a) $D_s D^*$ Monte Carlo,
          (b) $D_s^* D^*$ Monte Carlo, 
          (c) $D_s^{(*)} D^{**}$ Monte Carlo, 
          (d) $B\bar{B}$ background Monte Carlo, 
          (e) continuum data.
          The signals display a characteristic narrow peak, while the
          backgrounds are relatively broad. 
          }
\label{alpha_mc} 
\end{center} \end{figure}

In the case of the signal $B^0 \rightarrow D_s^+ D^{*-}$ mode, $\alpha$ is
small, $\alpha_0$ is as small or smaller, and the {\it difference} between
$\alpha$ and
$\alpha_0$ is very small.  Since the background is relatively isotropic in
$\cos{\alpha}$, it is more convenient to work with the cosines of the angles, 
where it is found that the difference $\cos{\alpha_0}-\cos{\alpha}$ peaks 
sharply for signal at small values.   Signal Monte Carlo distributions are
shown in Figure~\ref{alpha_mc} over the range $(-0.04, 2.00)$ for 
$D_s^+ D^{*-}$, $D_s^{*+} D^{*-}$, and $D_s^{(*)+} D^{**}$.  Two
backgrounds are also shown in the figure: the $B\bar{B}$ background, from
simulated non-signal $B$ meson events, and continuum background, from
$e^+ e^- \rightarrow c\bar{c}$, $s\bar{s}$, $u\bar{u}$, or $d\bar{d}$.  The
three signals display sharp peaking in $\cos{\alpha_0}-\cos{\alpha}$, where
the $D_s^* D^*$ and $D_s^{(*)} D^{**}$ peaks are measurably broader than the
$D_s D^*$.  In the $D_s^{*+} D^{*-}$ case, this broadness results from 
the random nudge given the $D_s^+$ by the $\gamma/\pi^0$ in the 
$D_s^{*+} \rightarrow D_s^+ \gamma / D_s^+ \pi^0$ transition, causing the
two particles to be not quite so back-to-back.  For the $D_s^{(*)+} D^{**}$,
the broad peak comes from the thrust given the $D^{*-}$ from the
unreconstructed $\pi^+$ produced during the intermediate
$\bar{D}^{**0} \rightarrow D^{*-} \pi^+$
decay.  It should be noted that no sharp peaking occurs in either
background where the $D_s^+$ and $\pi_s^-$ are nearly uncorrelated,
though some hint of a peak is exhibited due to kinematic correlations.  

It is seen from Figure~\ref{alpha_mc} that 
$\cos{\alpha_0}-\cos{\alpha}$ occasionally drifts below zero.  This is 
due to detector resolution effects that distort the quantities used to 
calculate $\cos{\theta_1}$ and $\cos{\theta_2}$. 

% have no cause to be back-to-back
% with each other for random events.  Some hint of a peak can, however,
% will be noted in the background at small values: since three charged tracks
% are taken to reconstruct the $D_s^+$ while only one is necessary for the
% $\pi_s^-$, there simply exist more charged tracks in the direction opposite
% the $D_s^+$.  
 
\subsection{Fit of the Data}

\begin{figure}[htbp] \begin{center}
\epsfig{figure=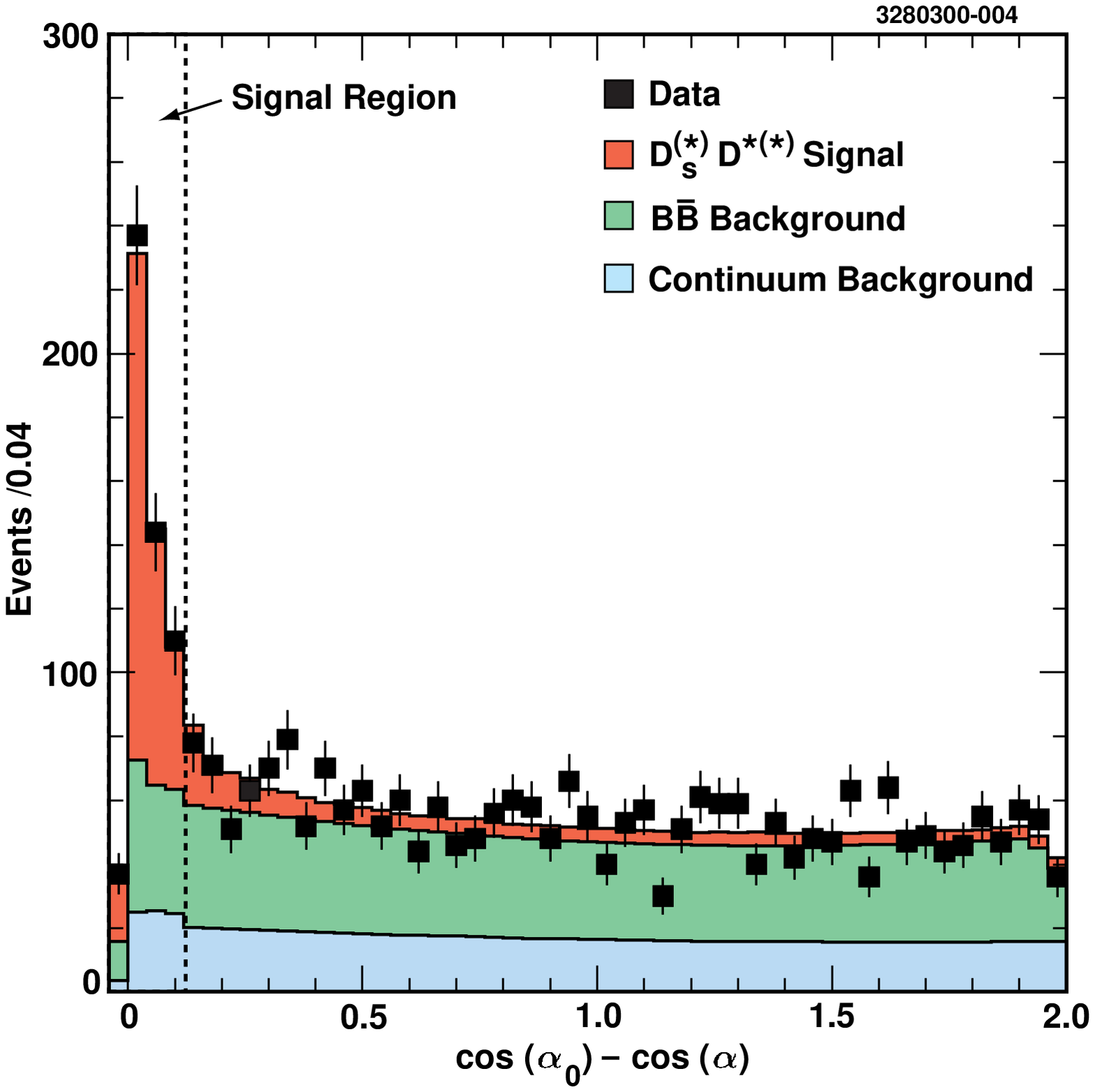, width=5.00in}
\caption{ The fitted $\cos{\alpha_0}-\cos{\alpha}$ data distribution. 
          The fit is broken down into three components: $D_s^{(*)} D^{*(*)}$
          signal, $B\bar{B}$ background, and continuum background. 
          The signal region is
          $-0.04 \leq \cos{\alpha_0}-\cos{\alpha} \leq 0.12$.
          The continuum background is constrained by scaling the off-resonance
          background by the on/off-resonance ratio.  
         }
\label{1d_fit}
\end{center} \end{figure}

A sharp signal peak in the $\cos{\alpha_0}-\cos{\alpha}$ distribution of the
CLEO on-resonance data previously described in Section II, superimposed on
a relatively flat background,  
is seen in Figure~\ref{1d_fit}.  The figure also shows a binned 
maximum-likelihood fit to this data
consisting of three components: $D_s^{(*)} D^{*(*)}$ signal, $B\bar{B}$
background, and continuum background.  The $D_s^{(*)} D^{*(*)}$ signal and
$B\bar{B}$ background components are allowed to float, while the continuum
level is fixed by scaling the off-resonance
background by the on/off-resonance ratio.  The $D_s^{(*)} D^{*(*)}$ 
component is a weighted combination of $D_s D^*$, $D_s^* D^*$, and
$D_s^{(*)} D^{**}$ signals, as the signal distribution shapes in
$\cos{\alpha_0}-\cos{\alpha}$ are too similar for meaningful separation. 
The signal is concentrated in the relatively 
small region $-0.04 \le \cos{\alpha_0}-\cos{\alpha} \le 0.12$, where the 
data contains 528 events.  Table~\ref{1d_results} lists fit results
of the three components for this signal region.  The errors
listed are statistical.  The subsequent analysis of the relative
$D_s^{(*)} D^{*(*)}$ rates and polarizations is confined to the signal region:
\begin{equation}
-0.04 \le \cos{\alpha_0}-\cos{\alpha} \le 0.12 
\end{equation}
\begin{table}[htb]
\caption{ Fit results for the data signal region $-0.04 \le 
          \cos{\alpha_0}-\cos{\alpha} \le 0.12$, containing 528 events. 
          Errors are statistical.  
         } 
\begin{center}
\begin{tabular}{ll}
Mode                        & Number of events \\ \hline 
$D_s^{(*)} D^{*(*)}$ Signal & $314.0 \pm 24.0$ \\ 
$B\bar{B}$ Background       & $138.9 \pm 5.2$  \\ 
Continuum                   & $74.8$ (constrained) \\
\end{tabular}
\end{center}
\label{1d_results}
\end{table} 

\section{SEPARATION OF $D_s D^*$, $D_s^* D^*$, AND $D_s^{(*)} D^{**}$ SIGNALS. 
         MEASUREMENT OF $D_s^* D^*$ POLARIZATION} 
\label{2d_discuss}

\subsection{Definition of the Two-Dimensional $p_{D_s}$ vs $\cos{\theta_{\pi}}$
            Parameter Space}

\begin{figure}[htbp] \begin{center}
\epsfig{figure=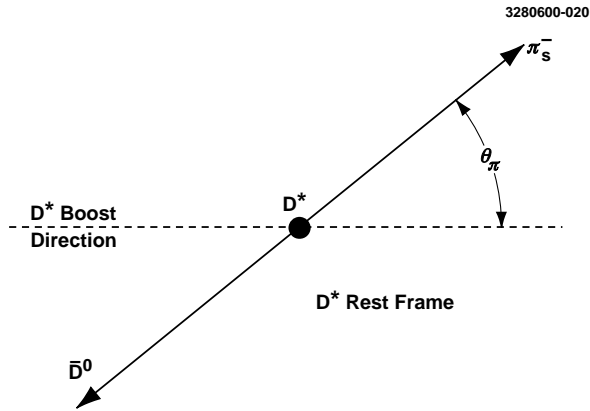, width=3.50in}
\caption{ Defining $\cos{\theta_{\pi}}$: the decay angle of the $\pi_s$
          as measured in the $D^*$ rest frame. 
         }
\label{theta_pi_def}
\end{center} \end{figure}

Once the background levels have been determined,  the signal modes may be
separated from one another.  This separation is effected by constructing a
two-dimensional parameter space, where each signal carries a distinctive
shape.  Two variables are required, of which the first is the magnitude of
$D_s^+$ momentum $p_{D_s}$.  The kinematics of the $B \rightarrow D_s^{(*)} 
D^{*(*)}$ decays constrain the relevant $D_s$
momentum range to $1250\ {\rm MeV}/c \le p_{D_s} \le 1925\ {\rm MeV}/c$.  The
second variable of interest is the cosine of the
$\pi_s^-$ decay angle as expressed in the $D^{*-}$
frame---$\cos{\theta_{\pi}}$---where $\theta_{\pi}$ is a helicity angle, 
shown in Figure~\ref{theta_pi_def}. 
It is possible to calculate $\cos{\theta_{\pi}}$
from available event information, without reconstructing the $D^{*-}$:

\begin{equation}
\cos{\theta_{\pi}} = - \frac{\beta_{D^*}(E^*_{\pi_s}-E^*_{D^0})}
                            {2 p^*_{D^0}}
                     + \frac{p^2_{\pi_s}-p^2_{D^0}}
                       {2 \gamma^2_{D^*} \beta_{D^*} M_{D^*} p^*_{D^0}}
\end{equation}

Here the kinematics of the $B^0 \rightarrow D_s^+ D^{*-}$ mode have been
assumed.
The quantities $\gamma_{D^*}$, $\beta_{D^*}$, $p_{\pi_s}$, and $p_{D_s}$ are
expressed in the lab frame, while $E^*_{\pi_s}$, $E^*_{D^0}$, and $p^*_{D^0}$
are in the $D^{*-}$ frame,

\begin{equation}
E^*_{D^0} = \frac{M^2_{D^*}+M^2_{D^0}-M^2_{\pi_s}}{2 M_{D^*}}
\end{equation}

\begin{equation}
E^*_{\pi_s} = \frac{M^2_{D^*}-M^2_{D^0}+M^2_{\pi_s}}{2 M_{D^*}}. 
\end{equation}

In the $B^0 \rightarrow D_s^+ D^{*-}$ mode, the $D^{*-}$ is produced in a 
$(J,J_z) = (1,0)$ state, and conservation of helicity distributes the
$\pi_s^-$ as $\cos^2{\theta_{\pi}}$.  Imperfect detector resolution smears
the shape.
For the case of longitudinally polarized $D_s^{*+} D^{*-}$ from $B^0$
decays,
the $D^{*-}$ is also produced in a $(1,0)$ state.  However, the resulting
$\cos^2{\theta_{\pi}}$ shape is not centered at the origin, but is rather
shifted downwards.  This shift comes from the missing
$\gamma/\pi^0$ (where $D_s^{*+} \rightarrow D_s^+ \gamma / D_s^+ \pi^0$),
which was not taken into account in the calculation of $\cos{\theta_{\pi}}$.
Nevertheless the original $\cos^2{\theta_{\pi}}$ shape is well
preserved, centered at $-0.2$, and falls over the range $(-1.2,0.8)$. 
In the case of transversely polarized $D_s^{*+} D^{*-}$,
the $D^{*-}$ is produced in a $(1,1)$ or $(1,-1)$ state, and the resulting
$\pi_s^-$ produces a helicity distribution of $1-\cos^2{\theta_{\pi}}$, 
also centered at $-0.2$.
Finally, the three $D^{**0}$ states each produce the $\pi_s^-$ in their
unique helicity distributions: the $D_1(2420)^0$ decays as
$1 + 3 \cos^2{\theta_{\pi}}$, the $D_2^*(2460)^0$ decays as
$1-\cos^2{\theta_{\pi}}$, and the $D_1(j=1/2)^0$ decays isotropically. 
However, blending the three $D^{**0}$ states according to their production
ratios in $D_s^+ D^{**}$ and $D_s^{*+} D^{**}$ effectively washes out any 
characteristic helicity shape.  The resulting blended distribution is centered
at $-0.4$ and ranges over $(-1.4, 0.5)$, because of the missing
intermediate $\pi$ (from $D^{**} \rightarrow D^{*-} \pi$), which is not
accounted for in calculating $\cos{\theta_{\pi}}$.  
The limits of $\cos{\theta_{\pi}}$ relevant to this analysis are therefore
$-1.40 \le \cos{\theta_{\pi}} \le 1.05$.

\begin{figure}[htbp] \begin{center}
\epsfig{figure=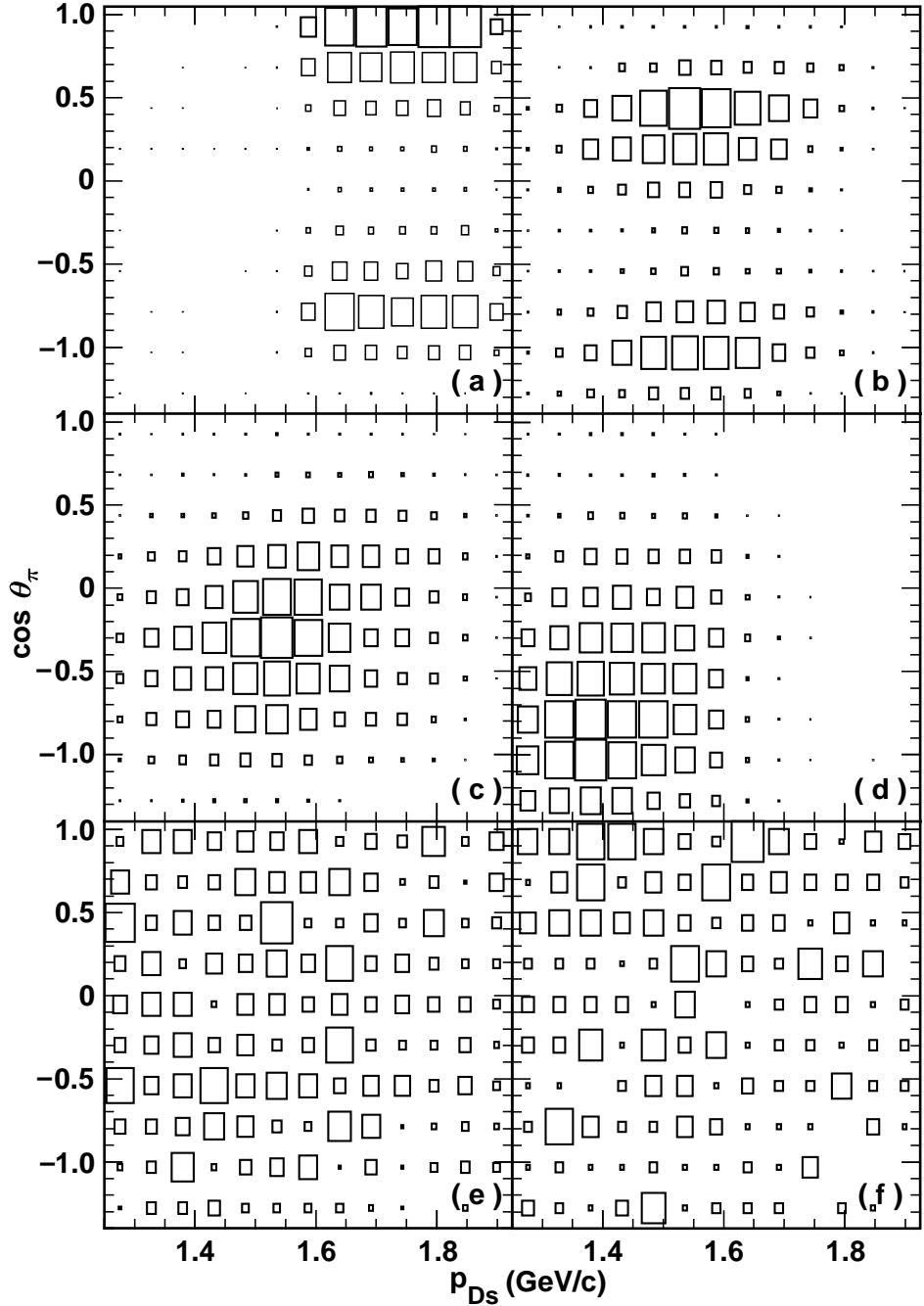, width=5.00in}
\caption{ $p_{D_s}$ vs $\cos{\theta_{\pi}}$ distributions
          for the four signal Monte Carlo and two background samples.   
          Shown are: 
         (a) $D_s D^*$ Monte Carlo,
         (b) Longitudinally Polarized $D_s^* D^*$ Monte Carlo,
         (c) Transversely Polarized $D_s^* D^*$ Monte Carlo,
         (d) $D_s^{(*)} D^{**}$ Monte Carlo,
         (e) $B\bar{B}$ background Monte Carlo,
         (f) continuum data.
         The box size is proportional to the number of candidates in the bin.
         }
\label{mc_2d}
\end{center} \end{figure}

\begin{figure}[htbp] \begin{center}
\epsfig{figure=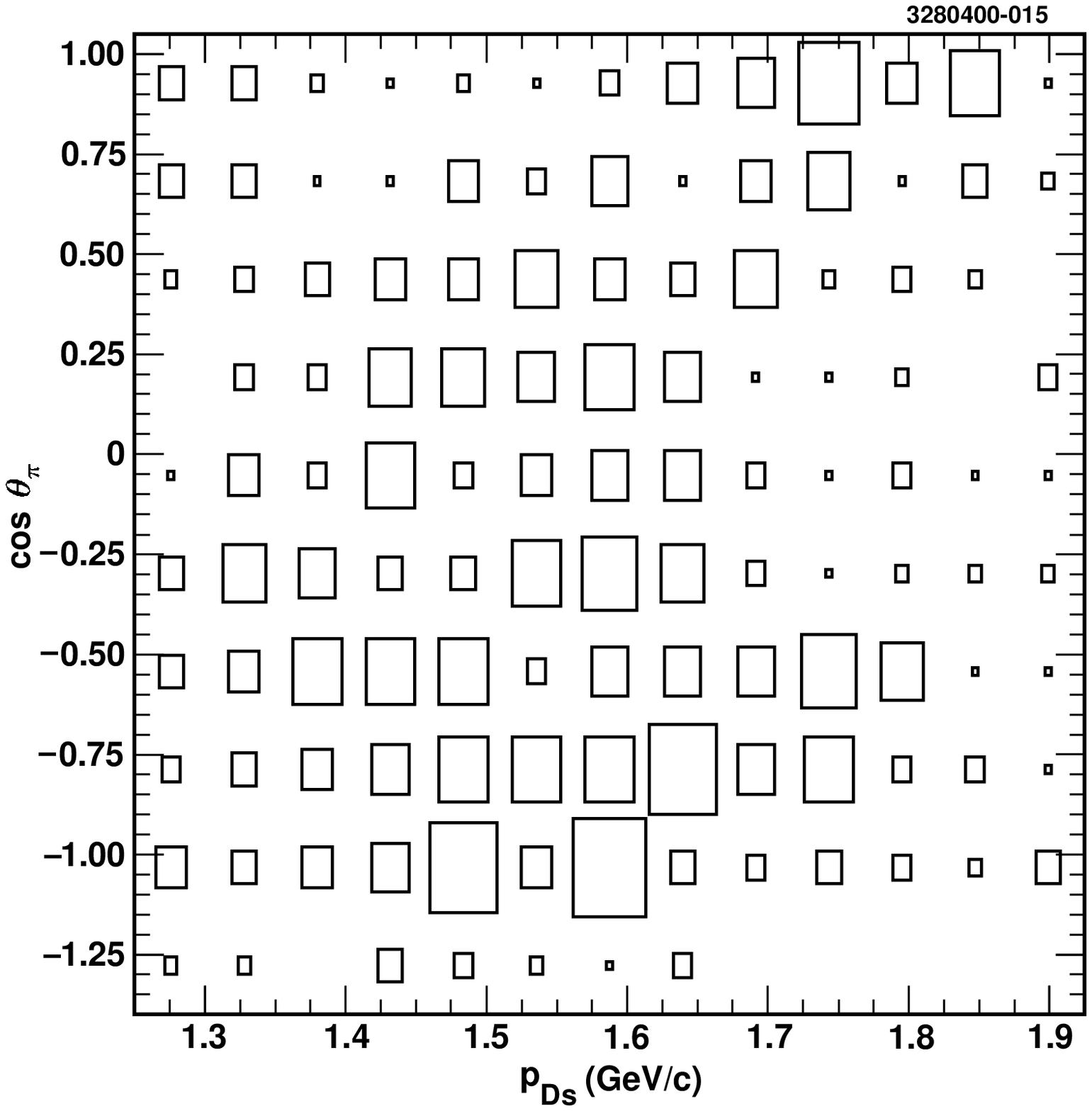, width=3.50in}
\caption{ $p_{D_s}$ vs $\cos{\theta_{\pi}}$ distributions for the CLEO 
          data.  The plot contains 528 events.
          The box size is proportional to the number of candidates in the bin.
        }
\label{data_2d}
\end{center} \end{figure}

The $p_{D_s}$ versus $\cos{\theta_{\pi}}$ two-dimensional distributions are
shown in Figure~\ref{mc_2d} for each of the four signals ($D_s D^*$,
Longitudinally Polarized $D_s^* D^*$, Transversely Polarized $D_s^* D^*$,
and $D_s^{(*)} D^{**}$) and two backgrounds ($B\bar{B}$ and continuum). 
Because the longitudinally
polarized and transversely polarized $D_s^* D^*$ produce markedly different
shapes in this two-dimensional distribution, they can be separated into two
components.  The two-dimensional on-resonance CLEO data distribution is shown
in Figure~\ref{data_2d}.

\subsection{The Fitted Data}

\begin{figure}[htbp] \begin{center}
\epsfig{figure=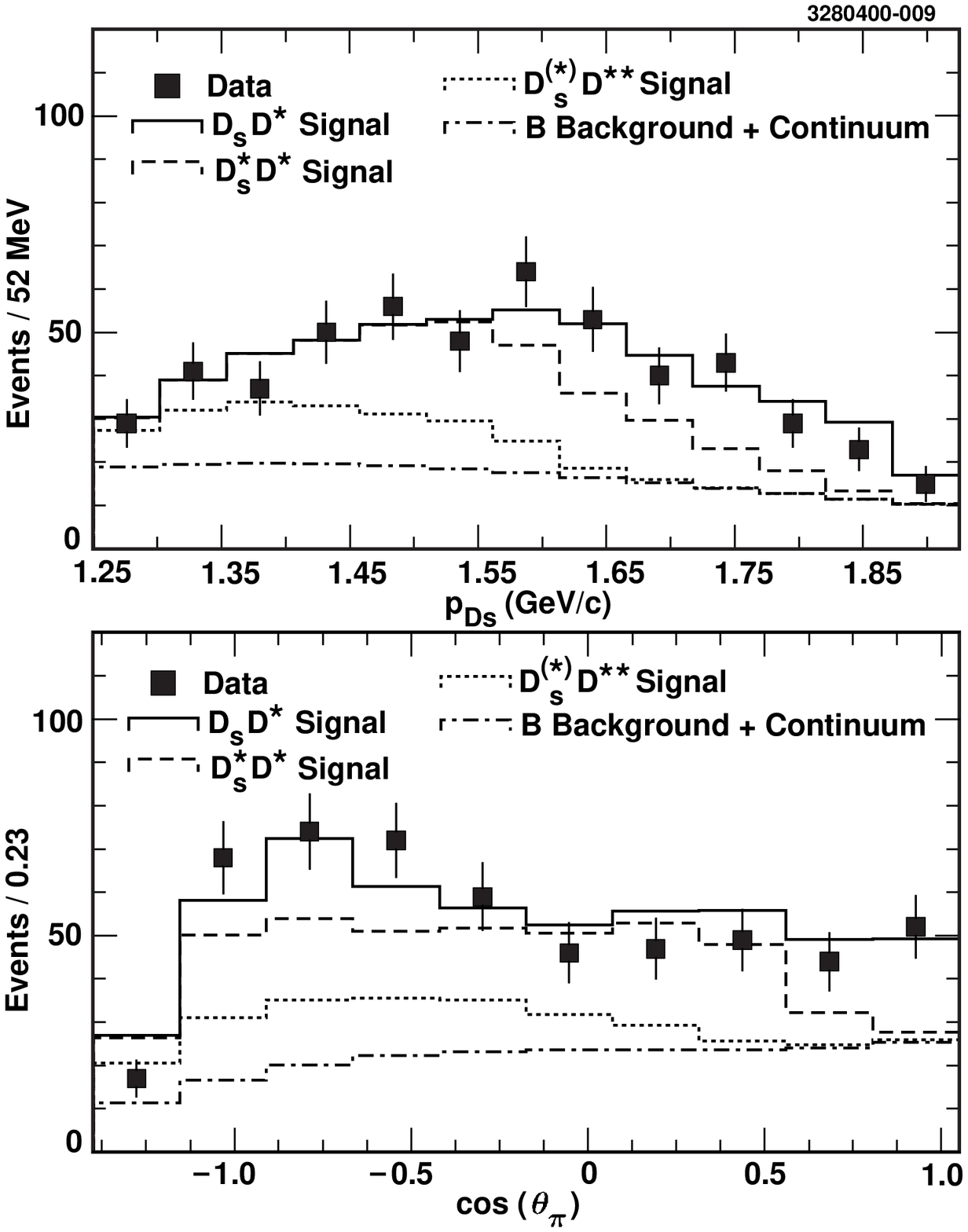, width=6.00in}
\caption{ The projections of the two-dimensional CLEO data and fit along the
          $p_{D_s^+}$
          axis {\it (top)} and $\cos{\theta_{\pi}}$ axis {\it (bottom)}. 
          The fit contains $D_s D^*$, $D_s^* D^*$, $D_s^{(*)} D^{**}$,
          and background components.  
          }
\label{2d_fit}
\end{center} \end{figure}

A two-dimensional binned maximum-likelihood fit is applied to the data.  
The $B\bar{B}$ and continuum backgrounds, whose levels were 
determined in the previous one-dimensional fit to the 
$\cos{\alpha_0}-\cos{\alpha}$ distribution, are fixed here, and their 
shapes are parameterized as products of Chebyshev polynomials.  
The two-dimensional Monte Carlo distributions are used for the
$D_s^{(*)} D^{*(*)}$ signals.
Four signal components are allowed to float: the number of $D_s D^*$, 
number of $D_s^* D^*$, number of $D_s^{(*)} D^{**}$, and the relative
longitudinal $D_s^* D^*$ polarization.  Converting the likelihood to 
a $\chi^2$-like quantity ($-2 \ln{\mathcal L}$), the resulting fit 
has a likelihood of 125.4 for 130 bins with 4 floating parameters.
Projections of data and fit along both the $p_{D_s}$ and $\cos{\theta_{\pi}}$
axes, broken down into signal and background components, are shown 
in Figure~\ref{2d_fit}.  
In Table~\ref{events}, the number of events resulting from the two-dimensional
maximum-likelihood fit are reported for each of the $D_s^{(*)+} D^{*(*)}$
modes, along with their statistical and systematic uncertainties, where
the systematics will be discussed in Section~\ref{sys_discuss}.
\begin{table}[htb]
\caption{ Fitted yield for each $D_s^{(*)+} D^{*(*)}$ mode.
          The first error is statistical and the second is systematic. 
          The $D^{**}$ is the sum of charged and neutral $D_1(2420)$,
          $D^*_2(2460)$, and $D_1(j=1/2)$ resonances. 
         } 
\begin{center}
\begin{tabular}{ll}
Mode                        & Fitted Yield \\ \hline 
$B^0 \rightarrow D_s^+ D^{*-}$       & $92.7  \pm 15.3 \pm  9.5$ \\ 
$B^0 \rightarrow D_s^{*+} D^{*-}$    & $149.2 \pm 30.4 \pm 20.9$ \\ 
$B   \rightarrow D_s^{(*)+} D^{**}$  & $81.6  \pm 23.3 \pm 15.3$ \\ 
\end{tabular}
\end{center}
\label{events}
\end{table} 

The relative longitudinal polarization of the $D_s^{*+} D^{*-}$ production
is measured to be:
\begin{equation}
\Gamma_L / \Gamma = (50.6 \pm 13.9 \pm 3.6)\%
\end{equation}
where the first error is statistical and the second systematic.  

\subsection{The Fitted Data With the $D_s^{(*)+} D^{**}$ Component Removed} 
\label{fit_wo_ddub}

\begin{figure}[htbp] \begin{center}
\epsfig{figure=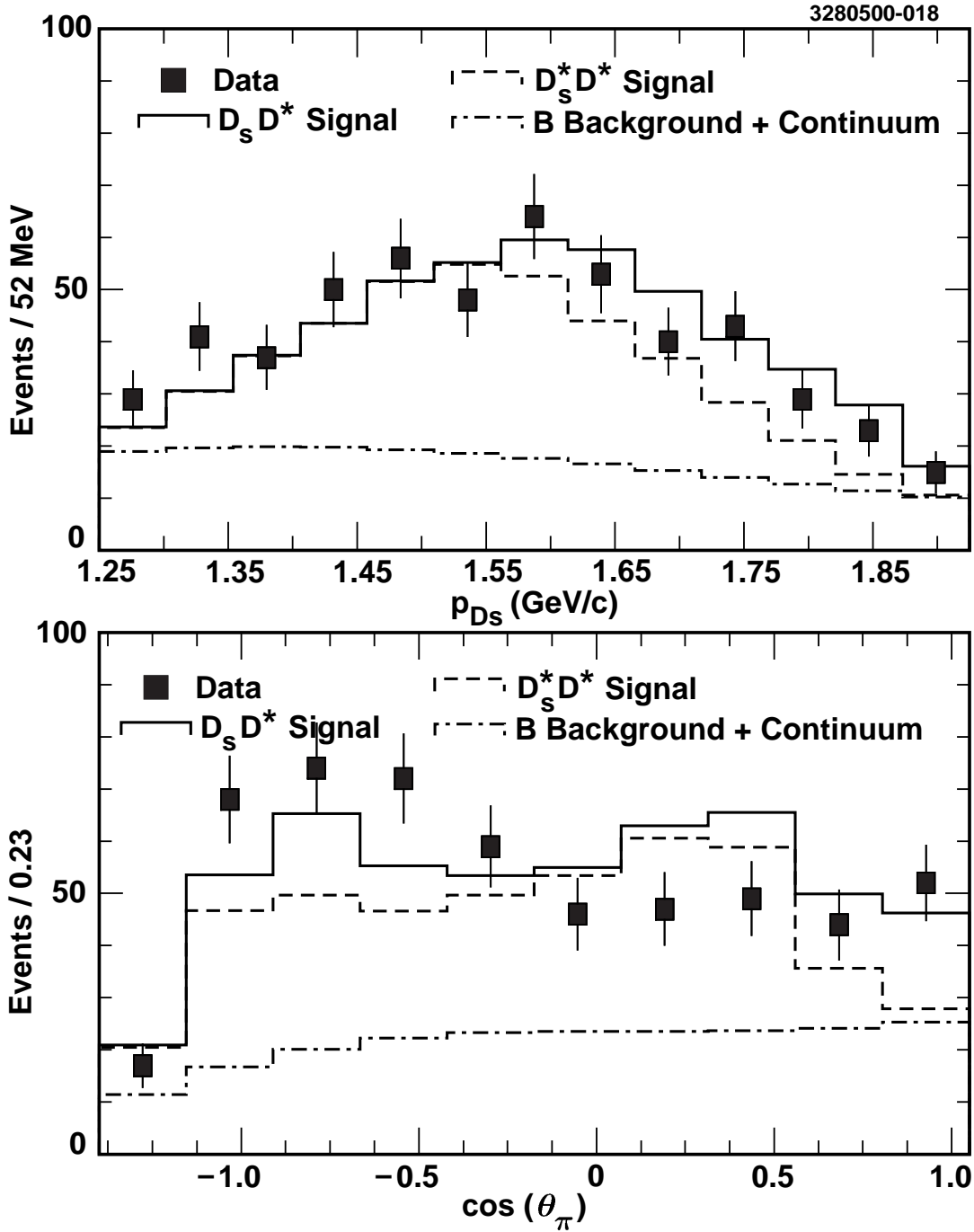, width=6.00in}
\caption{ Removing the $D_s^{(*)} D^{**}$ component from the two-dimensional
          fit to the CLEO data, where data and fit are projected along the
          $p_{D_s^+}$ axis {\it (top)} and $\cos{\theta_{\pi}}$ axis
          {\it (bottom)}.  The fit is split into $D_s D^*$, $D_s^* D^*$, 
          and background components.  The $p_{D_s}$ fit projection shape
          is shifted upwards relative to the data, and the ${\theta_{\pi}}$
          fit projection shape is systematically low over the region
          $-1.2 \le \cos{\theta_{\pi}} \le -0.2$ and systematically high over 
          the region $-0.2 \le \cos{\theta_{\pi}} \le 0.8$.  The likelihood
          is reduced by 3.8 standard deviations from the previous fit, which
          includes a $D_s^{(*)} D^{**}$ component.
          }
\label{2d_fit_noddub}
\end{center} \end{figure}

Because production of $B \rightarrow D_s^{(*)+} D^{**}$ has not been previously
observed, one might question its inclusion in the preceding fit.  A worthwhile
consistency check is to remove the $D_s^{(*)+} D^{**}$ from the set
of functions and repeat the two-dimensional fitting procedure.  The results,
without the $D_s^{(*)+} D^{**}$, are projected along the $p_{D_s}$ axis and
$\cos{\theta_{\pi}}$ axis in Figure~\ref{2d_fit_noddub}.  The $p_{D_s}$ fit
projection shape matches the data well, but is shifted systematically
upwards by about 50 MeV/$c$.  The $\cos{\theta_{\pi}}$ fit projection shape is
decidedly different from the data, since the projection is systematically
low over the region $-1.2 \le \cos{\theta_{\pi}} \le -0.2$ and
systematically high over the region $-0.2 \le \cos{\theta_{\pi}} \le 0.8$. 
The $\cos{\theta_{\pi}}$ fit projection also displays a pair of symmetric
peaks that are not reflected in the data.  This two-dimensional fit has
a likelihood of 139.9 for 130 bins with three floating parameters, and
since for this fitting procedure the likelihood follows closely the $chi^2$
behavior, this corresponds to a reduced significance of 3.8 standard
deviations.  A study of the data sideband regions
off the $\cos{\alpha_0}-\cos{\alpha}$ signal peak ({\it i.e.} where
$\cos{\alpha_0}-\cos{\alpha} \ge 0.12$) reveals an amount of
$D_s^{(*)+} D^{**}$ that is consistent with the
amount of $D_s^{(*)+} D^{**}$ observed in the signal region
$-0.04 \le \cos{\alpha_0}-\cos{\alpha} \le 0.12$.  Taking the factors all
together, we conclude that the 
data strongly indicate a substantial $D_s^{(*)+} D^{*(*)}$ component.  

\section{SYSTEMATIC UNCERTAINTIES}
\label{sys_discuss}

The single largest uncertainty in the analysis is the 25\% uncertainty in the
$D_s^+ \rightarrow \phi \pi^+$ branching fraction\cite{Ds-PhP}:
\begin{equation}
{\mathcal B}(D_s^+ \rightarrow \phi \pi^+) = (3.6 \pm 0.9)\% 
\end{equation}
This uncertainty is displayed separately from the other systematic
uncertainties, which
are listed in Table~\ref{systematics}.

\begin{table}[htb]
\caption{Systematic uncertainties in percent for
         $B \rightarrow D_s^{(*)+} D^{*(*)}$
         decays and $\Gamma_L/\Gamma$, the longitudinal polarization of 
         $D_s^{*+} D^{*-}$ .
         }
\begin{center}
\begin{tabular}{lllll}
 Source & $D_s^+ D^{*-}$ & $D_s^{*+} D^{*-}$ & $D_s^{(*)+}$ & $\Gamma_L/\Gamma$ \\ 
        &                &                   & $D^{**}$     &  \\ \hline
 $D_s$ Tracking                     & 3.0 & 3.0 & 3.0 & 3.0 \\
 $\pi_s$ Tracking                   & 5.0 & 5.0 & 5.0 & 5.0 \\
 Total Number of                    & 1.8 & 1.8 & 1.8 & --- \\
 $B\bar{B}$ Mesons                  &     &     &     &     \\ 
 Fit Normalization                  & 2.9 & 2.9 & 2.9 & --- \\
 Monte Carlo Statistics             & 1.0 & 1.0 & 1.0 & 1.0 \\
 Continuum Subtraction              & 3.7 & 3.7 & 3.7 & 1.8 \\
 $B\bar{B}$ Background              & 1.6 & 1.6 & 1.6 & 1.0 \\
 Subtraction                        &     &     &     &     \\
 Continuum Shape                    & 1.0 & 1.0 & 1.0 & 1.0 \\
 $B\bar{B}$ Background              & 1.0 & 1.0 & 1.0 & 1.0 \\
 Shape                              &     &     &     &     \\ 
 $D_1(j=1/2) : D_1(2420) :$         & 0.8 & 2.9 & 4.5 & 0.6 \\
 $D_2^*(2460)$ ratio                &     &     &     &     \\
 $D_s^+ D^{**} : D_s^{*+} D^{**}$ Ratio          & 2.4 & 9.8 & 14.8 & 3.2 \\ 
 Non-Resonant                       & 2.2 & 4.5 & 5.9 & 1.2 \\ 
 $D_s^{(*)+} D^{*-} \pi$ Production &     &     &     &     \\
\hline
 Total for $D_s^{(*)+} D^{*(*)}$ yield           & 9.9 & 13.8 & 18.5 & --- \\ 
 ${\mathcal B}(\phi \rightarrow K^+ \, K^-    )$ & 1.6  & 1.6 & 1.6 & ---\\
 ${\mathcal B}(D^{*-} \rightarrow \bar{D}^0 \, \pi_s^-)$ & 2.0  & 2.0 & 2.0 & ---\\ \hline
% ${\mathcal B}(D_s \rightarrow \phi \, \pi    )$ & 25.0 & 25.0 & 25.0  \\
% We took this out of the table to list it separately.  
 Total systematic uncertainty       & 10.2 & 14.0 & 18.7 & 7.3 \\ 
\end{tabular}
\end{center}
\label{systematics}
\end{table}

A 1\% systematic uncertainty in track finding and fitting efficiency is
estimated for each fast charged track, which for the $D_s^+$ add linearly to
a 3\% total.  The slow pion $\pi_s$ track finding and fitting uncertainty is
estimated at 5\%.  The uncertainty in the total number of $B\bar{B}$ meson
pairs introduces a systematic error of 1.8\%.  

The two-dimensional fit to the data estimates the total amount of 
$D_s^{(*)} D^{*(*)}$ signal at 323.2 events.  Since the previous
one-dimensional fit to the $\cos{\alpha_0}-\cos{\alpha}$ distribution,
summarized in Table~\ref{1d_results}, determined the level of 
$D_s^{(*)} D^{*(*)}$ signal at $314.0 \pm 24.0$, the two-dimensional
fit result overestimates the amount of signal by 9.2 events.  To test for a
systematic bias in the two-dimensional fitting procedure, fifty simulated
datasets were created and filled with $D_s^{(*)} D^{*(*)}$ signal Monte Carlo,
$B \bar{B}$ background, and continuum background according to the proportions
of Tables~\ref{1d_results} and~\ref{events}.  Following the procedure of
fixing both backgrounds and allowing all four signal components to float,
two-dimensional fits to these simulated datasets gave fifty estimates of
total $D_s^{(*)} D^{*(*)}$ signal.  The {\it difference} between the estimate
from each fit and the number of input $D_s^{(*)} D^{*(*)}$ events forms a
distribution centered at 1.1 with an rms of 5.3, consistent with zero
and indicative of an unbiased fitting procedure.   In the case of the 
fit to the real dataset, the additional 9.2 events 
differ from the expected total by an acceptable 1.7 standard deviations. 
In order that these events might be accounted for, a
systematic error of 2.9\% is introduced into the overall signal yield.
The polarization measurement is not affected by this systematic error 
in the fit normalization.

Forty thousand signal Monte Carlo events were generated for each of the
nine signal modes: $D_s^+ D^{*-}$, longitudinally polarized $D_s^{*+} D^{*-}$,
transversely polarized $D_s^{*+} D^{*-}$, $D_s^+ \bar{D}^{**0}$ (for each
of $\bar{D}_1(2420)^0$, $\bar{D}_2^*(2460)^0$, and $\bar{D}_1(j=1/2)^0$)
and $D_s^{*+} \bar{D}^{**0}$ (also for all 
three $\bar{D}^{**0}$ states).  To estimate statistical limitations, 
the signal samples were divided in half and the half-samples used to refit
the two-dimensional $p_{D_s^+}$ vs $\cos{\theta_{\pi}}$ data distribution.  The
resulting fits differ from the original by less than 1.0\%. 

An uncertainty is introduced by statistical fluctuations in the amount of
continuum background.  Varying the number of continuum background events by
one standard deviation ($\sigma$) affects the overall two-dimensional fit 
yields by a maximum of 3.7\% and the polarization by a maximum of 1.8\%.  
The uncertainty from statistical
fluctuations in the total number of $B\bar{B}$ background events is
anti-correlated with the continuum background.  This is the result of
highly similar background shapes in the one-dimensional fit to the
$\cos{\alpha_0}-\cos{\alpha}$ data distribution.  Refitting the two-dimensional 
data distribution with these fluctuations changes the yields by a maximum
of 1.6\%, and the polarization by a maximum of 1.0\%, where the small
uncertainty results from the anticorrelation. 

The two-dimensional continuum and background shapes are parameterized as
products of Chebyshev polynomials.  Varying the polynomial coefficients by
the parameterization errors and refitting the two-dimensional data distribution
changes the results by less than 1.0\% for either background.  

% $D_1(j=1/2) : D_1(2420) : D_2^*(2460)$ ratio 

In the two-dimensional fit to the data distribution, there is a single 
component containing $D_s^{(*)+} D^{**}$ signal.  The $D^{**}$ label
denotes the sum of three $L=1$ charm states: the $D_1(j=1/2)$, $D_1(2420)$,
and the $D_2^*(2460)$.  Each of these three states has a unique mass and
width, and produces a different pattern of $\pi_s^-$ helicities.  In building
the $D^{**}$ signal component, it is assumed that the $D^{**}$ production
rate from $B$ mesons is at a $D_1(j=1/2) : D_1(2420) : D_2^*(2460)$ ratio of
$2:1:6.7$, in accordance with the known $D_1(2420)$ and $D_2^*(2460)$
production rates in $B \rightarrow D^{**} \pi$\cite{D-Dub1} and the preliminary
evidence for the $D_1(j=1/2)$\cite{Wide-D}. 
To understand the systematic bias introduced by this choice of ratios,
the data was refit using widely varying ratios of ($2:1:3.3$, $2:1:13.5$,
$1:1:6.7$, $4:1:6.7$, $2:2:6.7$, and $2:0.5:6.7$).  This caused the
$D_s D^*$ yield to vary by 0.8\%, the $D^*_s D^*$ yield to vary by as
much as 2.9\%, the $D_s^{(*)} D^{**}$ yield to vary by 4.5\%, and the 
polarization to vary by 0.6\%.  

% $D_s^+ D^{**} : D_s^{*+} D^{**}$ Ratio  

The $D_s^+ D^{**} : D^{*+}_s D^{**}$ ratio in the $D_s^{(*)} D^{**}$ component
has been fixed {\it a priori} at $1:2$ in the two-dimensional fit.  The
assumption of this ratio follows from
the analogous modes $B \rightarrow D_s^{(*)} D^*$, where $D_s D^* : D^*_s D^*$
has been previously measured at $1:2$, a ratio confirmed by this
analysis.  The pseudoscalar/vector $D_s:D^*_s$ ratio in this spectator decay
implies that the same ratio should hold for the $D_s^{(*)} D^{**}$ case as
well.  All
the decays $B^0 \rightarrow D_s^{(*)+} \bar{D}^{*(*)}$ are spectator decays
described by a single Feynman diagram and differentiated only by the final
angular momentum states of the $c\bar{s}$ ($D_s^+$ or $D_s^{*+}$) and
$\bar{c}q$ ($D^{*}$ or $D^{**}$) quark pairs.  
To be particularly conservative, the $D_s D^{**} : D^*_s D^{**}$ ratio is
allowed to vary between $1:1$ and $1:4$.  Pseudoscalar/vector spin
considerations strongly suggest that the ratio be confined between these two
limits.  
Varying the $D_s^+ D^{**} : D^{*+}_s D^{**}$ ratio between $1:1$ and $1:4$ 
changes the fit results significantly, as the $D_s^+ D^{*-}$ varies by a
maximum of 2.4\%, the $D_s^{*+} D^{*-}$ by 9.8\%, the $D^{(*)+} D^{**}$
by 14.8\%, and the polarization by 3.2\%.  These errors are the second largest 
systematic uncertainty, 
after the $D_s^+ \rightarrow \phi \pi^+$ branching fraction uncertainty.  

% Non-Resonant $D_s^{(*)+} D^{*-} \pi$ Production 

There exists the possibility that significant non-resonant 
$B \rightarrow D_s^{(*)+} D^* \pi$ production could contribute to the data
sample.  The three-body $B \rightarrow D_s^{(*)+} D^* \pi$ decay peaks nearly
as strongly as resonant signal in $\cos{\alpha_0}-\cos{\alpha}$.  
%%%%%%%%%%%%%%%%%%%%%%%%%%%%%%%%%%%%%%%%%%%%%%%%%%
% FIX YOUR $&#@$*!! NONRESONANT DISCUSSION HERE!!
%%%%%%%%%%%%%%%%%%%%%%%%%%%%%%%%%%%%%%%%%%%%%%%%%%
While no measurements of the $B \rightarrow D_s^{(*)+} D^{*-} \pi$
non-resonant production have been made, an analogy can be drawn to
non-resonant production of $B \rightarrow D^{*-} \pi \ell \nu$.
ALEPH has measured the inclusive branching fraction
$B \rightarrow D^{*-} \pi \ell^- \nu$ at $(1.25 \pm 0.25)\%$, and the
product of exclusive branching fractions
${\mathcal B}(\bar{B} \rightarrow D_1(2420) \ell^- \nu)
{\mathcal B}(D_1(2420)^0 \rightarrow D^{*} \pi) =
(0.52 \pm 0.17)\% $\cite{Dst-l-nu2}.
ALEPH has also placed an upper limit on the $D_2^*(2460)$ branching
fraction at ${\mathcal B}(\bar{B} \rightarrow D_2^*(2460) \ell^- \nu)
{\mathcal B} (D_2^*(2460) \rightarrow D^{*} \pi) < 0.39\% $\cite{Dst-l-nu2}. 
Although there is no measurement of the mode
$\bar{B} \rightarrow D_1(j=1/2) \ell^- \nu$, 
recent observations at CLEO of the related mode
$B^+ \rightarrow \bar{D}^{**0} \pi^+$ report
${\mathcal B}(B \rightarrow \bar{D}_1(2420)^0 \pi^+) \approx
2/3 {\mathcal B}(B^+ \rightarrow \bar{D}_1(j=1/2)^0 \pi^+) \approx
2/3 {\mathcal B}(B^+ \rightarrow \bar{D}_2^*(2460)^0 \pi^+)$\cite{Wide-D}.  
Assuming that ${\mathcal B}(\bar{D}_1(2420)^0 \rightarrow D^{*-} \pi^+) = 
{\mathcal B}(\bar{D}_1(j=1/2)^0 \rightarrow D^{*-} \pi^+) = 2/3$ and
${\mathcal B}(\bar{D}_2^*(2460)^0 \rightarrow D^{*-} \pi^+) = 1/5$, and
assuming that these relative $D^{**} \pi$ ratios hold in the semileptonic
case, nearly all of the inclusive $B \rightarrow D^{*-} \pi \ell^- \nu$
will be accounted for by resonant $B \rightarrow D^{**} \ell^- \nu$.  This
would leave only a small nonresonant component.  
% However, due to the 
% high $q^2$ of this decay, the non-resonant $D_s^{(*)} D^* \pi$ production rate
% is suppressed relative to non-resonant $B \rightarrow D^{*} X \ell \nu$.
Thus a conservative upper
limit is that non-resonant $B \rightarrow D_s^{(*)+} D^{*-} \pi$ could be
as large as 40\% of the resonant $B \rightarrow D_s^{(*)+} D^{**}$ branching 
fraction.  Three $B \rightarrow D_s^{(*)} D^{**}$ $+$ (Non-Resonant) 
samples were created:
one that contained
60\% pure $B \rightarrow D^{(*)+} D^{**}$ with
30\% non-resonant $B \rightarrow D_s^+ D^{*-} \pi$ and 
10\% $B \rightarrow D^{*+}_s D^{*-} \pi$,
one that contained
60\% pure $B \rightarrow D_s^{(*)+} D^{**}$ with
10\% non-resonant $B \rightarrow D_s^+ D^{*-} \pi$ and 
30\% non-resonant $B \rightarrow D^{*+}_s D^{*-} \pi$,
and one that contained 60\% pure 
$B \rightarrow D_s^{(*)+} D^{**}$ with 
20\% non-resonant $B \rightarrow D_s^+ D^{*-} \pi$ and
20\% non-resonant $B \rightarrow D^{*+}_s D^{*-} \pi$.  
Refitting the data distribution with
these $B \rightarrow D_s^{(*)+} D^{**}$ $+$ non-resonant $B \rightarrow 
D_s^{(*)+} D^{*-} \pi$ samples changes the results by 2.2\% for the
$B \rightarrow D_s^+ D^{*-}$ case, by 4.5\% for the
$B \rightarrow D^{*+}_s D^{*-}$, by 5.9\% for the
$B \rightarrow D_s^{(*)+} D^{**}$, and by 1.2\% for the polarization.  
These are the systematic errors listed in Table~\ref{systematics}. 
Should it be the case that
by 60\% of resonant $B \rightarrow D_s^{(*)+} D^{**}$ branching fraction
be non-resonant, the systematic errors increase would increase to 3.0\%
for the $B \rightarrow D_s^+ D^{*-}$, to 6.3\% for the
$B \rightarrow D^{*+}_s D^{*-}$, to 8.1\% for the
$B \rightarrow D_s^{(*)+} D^{**}$, and to 1.8\% for the polarization. 
It should be noted that other non-resonant
modes, such as $B \rightarrow D_s^+ D^{*-} \pi \pi$, produce the $D_s^+$ in
a momentum range that is almost entirely below the lower limit of 1250
MeV/$c$, excluding these modes from this analysis. 

The 1998 PDG values for the $\phi$ and $D^{*-}$ branching fractions
are ${\mathcal B}(\phi \rightarrow K^+ K^-) = (49.1 \pm 0.8)\%$ and 
${\mathcal B}(D^{*-} \rightarrow \bar{D}^0 \pi_s^-) =
(68.3 \pm 1.4)\%$\cite{PDG}. 
These introduce systematic errors of 1.6\% and 2.0\%, respectively, into the 
extraction of the $D_s^{(*)+} D^{*(*)}$ branching fractions.

It is assumed in measuring the longitudinal and transverse 
$D_s^{*+} D^{*-}$ polarizations that these final states are independent
of one another.  In actuality there exists, in the differential decay rate,
an interference term between the longitudinal and transverse states
that is proportional to the azimuthal angle 
between the planes of the $D_s^{*+} \rightarrow D_s^+ \gamma$ and
$D^{*-} \rightarrow \bar{D}^0 \pi^-$ decays.  This interference vanishes
in the integral over the azimuth, and introduces no systematic error 
into the analysis.

\section{FACTORIZATION AND PREDICTION OF POLARIZATIONS}
\label{factorization}

The factorization assumption, when expressed in the framework of Heavy Quark 
Effective Theory (HQET) and extrapolating from the form factors measured by
the semileptonic
$B$ decays $B \rightarrow D^* \ell \nu$, allows accurate estimate of the
hadronic decay rates for the modes $B \rightarrow D^{(*)} \pi$, $D^{(*)}
\rho$, $D_s^{(*)} D^{(*)}$, and
$D^{(*)} D^{(*)}$\cite{Theory1,Big-B,Neubert1,Neubert2}. 
Additionally, factorization, HQET, and the semileptonic decays,
predict the relative polarization of the vector-vector hadron
products for $B \rightarrow D^{*-} X$ decays, such as $B \rightarrow
D^{*-} \rho$ and $B^0 \rightarrow D_s^{*+} D^{*-}$\cite{Richman,Dst-l-nu}.  
 
\begin{figure}[htbp] \begin{center}
\epsfig{figure=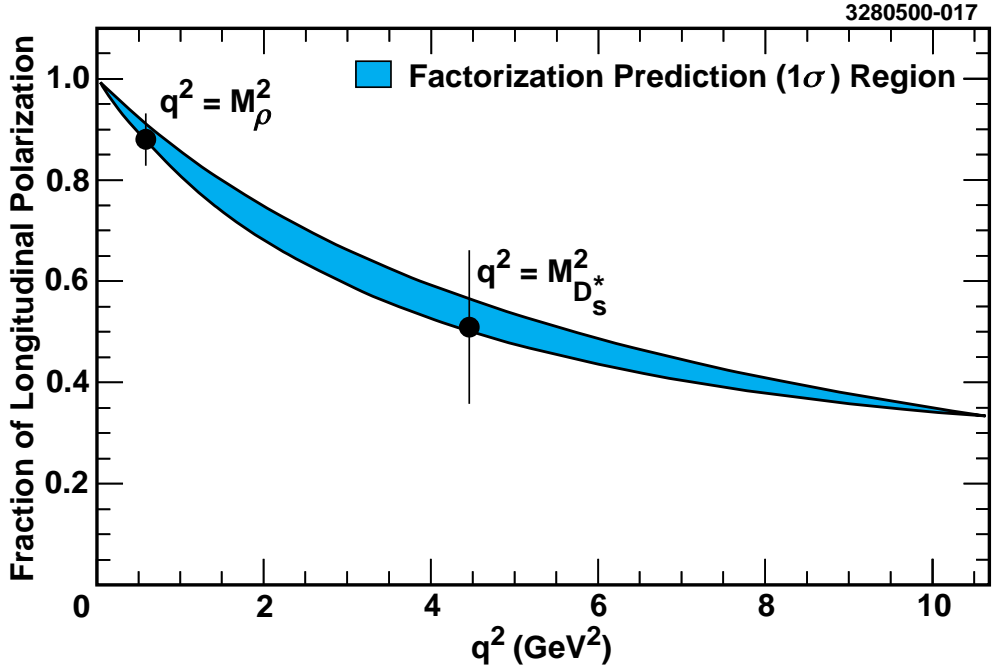, width=5.50in}
\caption{ Relative fraction of longitudinal polarization in vector-vector
          $B \rightarrow D^{*-} X$ decays as a function of $q^2$, where
          $q^2 = M^2_X$, and $X$ is a vector meson.  Shown are the 1998
          measurement of $B \rightarrow D^{*-} \rho$, and the 
          $B^0 \rightarrow D_s^{*+} D^{*-}$ polarization measured here for
          the first time.  The shaded region represents the prediction using
          factorization and Heavy Quark Effective Theory, and extrapolating
          from the semileptonic $B \rightarrow D^* \ell \nu$ form factor
          results.  The contour is one standard deviation ($\sigma$).
          }
\label{factorize}
\end{center} \end{figure}

We observe a longitudinal polarization in $B^0 \rightarrow D_s^{*+} D^{*-}$ of
$\Gamma_L/\Gamma =(50.6 \pm 13.9 \pm 3.6)\%$ for $q^2 = M^2_{D_s^*}$, 
where the first error is statistical and the second systematic.  The 
observation is consistent with the prediction of $(53.5 \pm 3.3)\%$ from 
factorization, HQET, and the semileptonic form factor
measurements\cite{Richman}.
The same combination also predicts in $B \rightarrow D^{*-} \rho$ a longitudinal
polarization of $\Gamma_L/\Gamma = (89.5 \pm 1.9)\%$ at $q^2 = M_{\rho}^2$,
which compares favorably with the most recent measurement
of $(87.8 \pm 5.3)\%$\cite{B-DstRho}.  Finally, predictions are also made
that at low $q^2$ the longitudinal polarization will be nearly 100\%, and
at $q^2 = q^2_{max}$ decreases to 33\%\cite{Theory1}.  
Longitudinal polarization as a function of $q^2$ is plotted in
Figure~\ref{factorize} for the factorization prediction and 
compared with the $D^{*-} \rho$ and $D_s^{*+} D^{*-}$ measurements.
The agreement is excellent, confirming the validity of the factorization
assumption and HQET in extrapolating the semileptonic form factor results
for regions of high $q^2$.  The polarization of the semileptonic decays
remains unobserved\cite{Dst-l-nu}.

Another vector-vector hadronic $B$ decay mode which may further test
the factorization assumption at high $q^2$ is $B^0 \rightarrow D^{*+}
D^{*-}$.  This decay is Cabibbo-suppressed, and a polarization
measurement will require higher statistics than those provided by
present experiments\cite{B-DstDst}.  Future experiments will also
reduce the errors of the $D^{*-} \rho$ and $D_s^{*+} D^{*-}$
measurements.

\section{SUMMARY AND CONCLUSIONS} 
\label{conclusion}

Removing the $D_s^{(*)+} D^{**}$ signal component from the two-dimensional
fit reduces the likelihood by $3.8 \sigma$, and the resulting projections 
along both the $p_{D_s}$ and $\cos{\theta_{\pi}}$ axes are systematically 
different from the data as discussed in Section~\ref{fit_wo_ddub}.   
Furthermore, a level of $D_s^{(*)+} D^{**}$ is observed in the data 
sideband regions of $\cos{\alpha_0}-\cos{\alpha}$ consistent with that
seen in the signal region.  
We conclude that the data support first evidence for
$B \rightarrow D_s^{(*)+} D^{**}$ decays.  
 
From the event yield of Table~\ref{events}, we can calculate the 
exclusive branching fractions $B^0 \rightarrow D_s^+ D^{*-}$,
$B^0 \rightarrow D_s^{*+} D^{*-}$, and 
$B^+ \rightarrow D_s^{(*)+} \bar{D}^{**0}$, where the $\bar{D}^{**0}$
is the sum of the $D_1(2420)^0$, $D^*_2(2460)^0$, and $D_1(j=1/2)^0$ 
states 
\begin{equation}
{\mathcal B} (B^0 \rightarrow D_s^+ \, D^{*-})  = 
    (1.10 \pm 0.18 \pm 0.11 \pm 0.28)\%,  
\end{equation}
\begin{equation}
{\mathcal B} (B^0 \rightarrow D_s^{*+} \, D^{*-})  = 
    (1.82 \pm 0.37 \pm 0.25 \pm 0.46)\%, 
\end{equation}
\begin{equation}
{ \mathcal B} (B^+ \rightarrow D_s^{(*)+} \, \bar{D}^{**0})  = 
    (2.73 \pm 0.78 \pm 0.51 \pm 0.68)\%.  
\end{equation}
The first error is statistical, the second systematic, and the third the
contribution from the uncertainty of the
$D_s^+ \rightarrow \phi \pi^+$ branching 
fraction.  These $B^0 \rightarrow D_s^{(*)+} D^{*-}$ branching
fractions supersede the previous CLEO measurements\cite{B-DsX}.
The extraction of the combined $D_s^{(*)+} \bar{D}^{**0}$
branching fraction is contingent on the assumption of Equation (7),
where the charged-$B$ decay rate, $B^+ \rightarrow D_s^{(*)+} \bar{D}^{**0}$,
is presumed equal to the neutral-$B$ decay rate, 
$B^0 \rightarrow D_s^{(*)+} D^{**-}$.  The extraction also requires some
presumption of the individual $D^{**} \rightarrow D^{*-} \pi$ rates, 
shown in Equations (1)--(6).  The assumptions follow from
conservation of isospin in the spectator $B$ decay of Figure~\ref{feynman}.
It is further assumed that the production rates of $B^+$ and $B^0$ in 
$\Upsilon(4S)$ decays are equal for all branching fraction measurements. 

The relative longitudinal $D_s^*$ polarization in
$B^0 \rightarrow D_s^{*+} D^{*-}$ is measured for the first time as: 
\begin{equation}
\frac{\Gamma_L}{\Gamma}(B^0 \rightarrow D_s^{*+} D^{*-}) =
(50.6 \pm 13.9 \pm 3.6)\%
\end{equation}
where the first error is statistical and the second systematic.  The
measurement is consistent with the recent factorization prediction
of $(53.5 \pm 3.3)\%$, confirming the validity of the factorization
assumption in the domain of relatively high $q^2$\cite{Richman}.

%%%%%%%%%%%%%%%%%%%%%%%%%%%%%%%%%%%%%%%%%%%%%%%%%%%%%%%%%%%%%%%%%%% 
% CURRENT ACKNOWLEDGMENTS go here...
\section*{ACKNOWLEDGMENTS} 
We gratefully acknowledge the effort of the CESR staff in providing us with
excellent luminosity and running conditions.
I.P.J. Shipsey thanks the NYI program of the NSF, 
M. Selen thanks the PFF program of the NSF, 
A.H. Mahmood thanks the Texas Advanced Research Program,
M. Selen and H. Yamamoto thank the OJI program of DOE, 
M. Selen and V. Sharma 
thank the A.P. Sloan Foundation, 
M. Selen and V. Sharma thank the Research Corporation, 
F. Blanc thanks the Swiss National Science Foundation, 
and H. Schwarthoff and E. von Toerne
thank the Alexander von Humboldt Stiftung for support.  
This work was supported by the National Science Foundation, the
U.S. Department of Energy, and the Natural Sciences and Engineering Research 
Council of Canada.
%%%%%%%%%%%%%%%%%%%%%%%%%%%%%%%%%%%%%%%%%%%%%%%%%%%%%%%%%%%%%%%%%%% 

\end{document}